\documentclass[twocolumn,tighten,times]{aastex631}
\usepackage{amsfonts}
\usepackage{mathrsfs}
\usepackage{bm}
\usepackage{lipsum}
\usepackage{physics}
\usepackage{multirow}
\usepackage{rotating}
\usepackage{graphicx} 
\usepackage{soul}
\usepackage{subfigure}
\usepackage{color}

\def\pp{\prime\prime}
\def\sunm{M_{\odot}}

\newcommand{\feii}{\ifmmode {\rm Fe\ II} \else Fe~{\sc ii}\fi}
\newcommand{\heii}{\ifmmode {\rm He\ II} \else He~{\sc ii}\fi}
\newcommand{\hei}{\ifmmode {\rm He\ I} \else He~{\sc i}\fi}
\newcommand{\oiii}{\ifmmode {\rm~[O\ III]} \else [O~{\sc iii}]\fi}
\newcommand{\nii}{\ifmmode {\rm~[N\ II]} \else [N~{\sc ii}]\fi}
\newcommand{\mgii}{\ifmmode {\rm~Mg\ II} \else Mg~{\sc ii}\fi}
\newcommand{\ciii}{\ifmmode {\rm~C\ III]} \else C~{\sc iii}]\fi}
\newcommand{\civ}{\ifmmode {\rm~C\ IV} \else C~{\sc iv}\fi}
\newcommand{\hb}{\rm{H$\beta$}}
\newcommand{\ha}{\rm{H$\alpha$}}
\newcommand{\hg}{\rm{H$\gamma$}}
\newcommand{\mbh}{\ifmmode {M_{\bullet}} \else $M_{\bullet}$\fi}
\newcommand{\dotm}{\ifmmode {\dot{\mathscr{M}}} \else $\dot{\mathscr{M}}$\fi}
\def\ergs{\rm erg\,s^{-1}}
\def\ergscm{\rm erg\,s^{-1}\,cm^{-2}}
\def\ergscma{\rm erg\,s^{-1}\,cm^{-2}\,\AA^{-1}}
\def\kms{\rm km\,s^{-1}}
\shorttitle{Reverberation Mapping of NGC~4151 }
\shortauthors{Li et al.}

\begin{document}

\title{\bf \large Velocity-resolved Reverberation Mapping of Changing-look Active Galactic Nucleus NGC~4151 During Outburst Stage: Evidence for Kinematics Evolution of Broad-line Region}

\correspondingauthor{Hai-Cheng Feng, H. T. Liu, $\&$ J. M. Bai}
\email{hcfeng@ynao.ac.cn, htliu@ynao.ac.cn, baijinming@ynao.ac.cn}

\author[0000-0003-3823-3419]{Sha-Sha Li}
\affiliation{Yunnan Observatories, Chinese Academy of Sciences, Kunming, 650011, Yunnan, People's Republic of China}
\affiliation{Key Laboratory for the Structure and Evolution of Celestial Objects, Chinese Academy of Sciences, Kunming 650011, Yunnan, People's Republic of China}

\author[0000-0002-1530-2680]{Hai-Cheng Feng}
\affiliation{Yunnan Observatories, Chinese Academy of Sciences, Kunming, 650011, Yunnan, People's Republic of China}
\affiliation{Key Laboratory for the Structure and Evolution of Celestial Objects, Chinese Academy of Sciences, Kunming 650011, Yunnan, People's Republic of China}
\affiliation{University of Chinese Academy of Sciences, Beijing 100049, People's Republic of China}

\author[0000-0002-2153-3688]{H. T. Liu}
\affiliation{Yunnan Observatories, Chinese Academy of Sciences, Kunming, 650011, Yunnan, People's Republic of China}
\affiliation{Key Laboratory for the Structure and Evolution of Celestial Objects, Chinese Academy of Sciences, Kunming 650011, Yunnan, People's Republic of China}
\affiliation{Center for Astronomical Mega-Science, Chinese Academy of Sciences, 20A Datun Road, Chaoyang District, Beijing 100012, People's Republic of China}

\author{J. M. Bai}
\affiliation{Yunnan Observatories, Chinese Academy of Sciences, Kunming, 650011, Yunnan, People's Republic of China}
\affiliation{Key Laboratory for the Structure and Evolution of Celestial Objects, Chinese Academy of Sciences, Kunming 650011, Yunnan, People's Republic of China}
\affiliation{Center for Astronomical Mega-Science, Chinese Academy of Sciences, 20A Datun Road, Chaoyang District, Beijing 100012, People's Republic of China}

\author[0000-0002-3490-4089]{Rui Li}
\affiliation{University of Chinese Academy of Sciences, Beijing 100049, People's Republic of China}

\author[0000-0002-2310-0982]{Kai-Xing Lu}
\affiliation{Yunnan Observatories, Chinese Academy of Sciences, Kunming, 650011, Yunnan, People's Republic of China}
\affiliation{Key Laboratory for the Structure and Evolution of Celestial Objects, Chinese Academy of Sciences, Kunming 650011, Yunnan, People's Republic of China}
\affiliation{Center for Astronomical Mega-Science, Chinese Academy of Sciences, 20A Datun Road, Chaoyang District, Beijing 100012, People's Republic of China}

\author[0000-0003-4156-3793]{Jian-Guo Wang}
\affiliation{Yunnan Observatories, Chinese Academy of Sciences, Kunming, 650011, Yunnan, People's Republic of China}
\affiliation{Key Laboratory for the Structure and Evolution of Celestial Objects, Chinese Academy of Sciences, Kunming 650011, Yunnan, People's Republic of China}
\affiliation{Center for Astronomical Mega-Science, Chinese Academy of Sciences, 20A Datun Road, Chaoyang District, Beijing 100012, People's Republic of China}

\author{Ying-Ke Huang}
\affiliation{Qian Xuesen Laboratory of Space Technology, China Academy of Space Technology, 104 Youyi Road, Haidian District, Beijing, 100094, People's Republic of China}
\affiliation{Multi-disciplinary Research Division, Institute of High Energy Physics, Chinese Academy of Sciences, 19B Yuquan Road, Beijing 100049, People's Republic of China}

\author{Zhi-Xiang Zhang}
\affil{Department of Astronomy, Xiamen University, Xiamen, Fujian 361005, People’s Republic of China}

\begin{abstract}
Changing-look active galactic nucleus NGC~4151, which has attracted a lot of attention, is undergoing the second dramatic outburst stage in its evolutionary history. To investigate the geometry and kinematics of the broad-line region (BLR), and measure the mass of supermassive black hole in NGC~4151, we perform a seven-month photometric and spectroscopic monitoring program in 2020--2021, using the 2.4 m telescope at Lijiang Observatory. We successfully measure the time lags of the responses from broad \ha, \hb, \hg, \hei, and \heii\ emission lines to continuum variation, which are $7.63_{-2.62}^{+1.85}$, $6.21_{-1.13}^{+1.41}$, $5.67_{-1.94}^{+1.65}$, $1.59_{-1.11}^{+0.86}$, and $0.46_{-1.06}^{+1.22}$ days, respectively, following radial stratification. The ratios of time lags among these lines are $1.23 : 1.00 : 0.91 : 0.26 : 0.07$. We find that the continuum lag between the ultraviolet and optical bands can significantly affect the lag measurements of \hei\ and \heii. Virial and infalling gas motions coexist in this campaign, which is different from previous results, implying the evolutionary kinematics of BLR. Based on our measurements and previous ones in the literature, we confirm that the BLR of NGC~4151 is basically virialized. Finally, we compute the black hole mass through multiple lines, and the measurement from \hb\ to be $ 3.94_{-0.72}^{+0.90} \times 10^7 M_{\odot}$, which is consistent with previous results. The corresponding accretion rate is $0.02_{-0.01}^{+0.01} L_{\rm Edd} c^{-2}$, implying a sub-Eddington accretor. 

\end{abstract}


\section{Introduction} \label{sec:intro}
Broad emission lines, one of the most prominent features in active galactic nuclei (AGNs), are emitted from the broad-line region (BLR). The BLR is one of the main components in the interior part of AGNs located between the accretion disk and dust structure \citep{Antonucci1993}. It is generally believed that motions of BLR clouds are dominated by gravity of central supermassive black hole (SMBH), while some researchers argued that the radiation pressure from the accretion disk can also contribute to non-negligible effects. Although broad emission lines allow to study the geometry and kinematics of the BLR, and measure the mass of SMBH, the BLR is still not yet fully understood. The BLR is too small to be directly spatial resolved by most telescopes. To date, only three AGNs have been resolved using the infrared interferometry \citep[e.g.,][]{Gravity2018}, therefore, other approaches are urgently needed. 

Among them, reverberation mapping (RM) is an effective way of investigating the BLR and SMBH, which subtly uses the time resolution of the telescope to substitute the spatial resolution \citep[e.g.,][]{Blandford1982,Cackett2021}. In principle, the continuum radiated from the accretion disk acts as ionizing photons to ionize the gas of the BLR, and this photoionization process generates the emission lines that are broadened by the BLR cloud motion into the corresponding broad emission lines. The characteristic size of the BLR can be inferred from monitoring the responses of broad emission lines to continuum variations and measuring the time delays between broad emission-line and continuum light curves. When combining the size of the BLR with broad emission-line velocity, and using the virial relationship, one can estimate the mass of SMBH. Over the past four decades, the RM method has measured the black hole masses of $\sim$ 100 AGNs, and has become a standard tool for measuring the masses of SMBHs in AGNs \citep[e.g.,][]{kaspi2000, peterson2004, bentz2009, barth2013, du2014, Fausnaugh2017, derosa2018, lu2019,zhang2019, hu2021, bao2022}. Additionally, the application of RM to the accretion disk and torus can measure the size of disk \citep[e.g.,][]{edelson2017, Cackett2020, Fian2022}, and determine dust sublimation radius \citep[e.g.,][]{Koshida2014, Oknyansky2014, lyu2021, lyu2022}. 

Here, the geometry and kinematics of BLR can be inferred from velocity-resolved RM analysis utilizing high-quality spectral data. The details are derived by measuring time lag as a function of line-of-sight velocity for the broad emission line. Currently, more than 30 AGNs have been measured using this method, and different kinematic characteristics of their BLR, such as rotation, inflow, and outflow, have been obtained (e.g., \citealt{du2016, Williams2018, Brotherton2020, Horne2021, lu2021, U2022, Villafana2022}). And a few well-studied AGNs were measured multiple times for velocity-resolved results in different periods. For example, \citet{xiao2018} recovered velocity-delay maps of NGC 5548 from multi-year data and found that the kinematics of the BLR varied between inflow and virial. Then, \citet{hu2020b} detected that the kinematics of the BLR in PG 2130+099 changed from virialized motion to inflow, and the timescale of change was less than one year. However, some other AGNs, e.g., NGC 3516, showed nearly consistent velocity-resolved signature over a decade \citep{denney2010, feng2021a}, indicating the complicated evolution in BLR. Therefore, to investigate the kinematic evolution of BLR, multiple velocity-resolved time delay measurements are necessary.

To achieve the above-mentioned purpose of studying the kinematic evolution of BLR, NGC~4151 is selected for RM monitoring in this work. As one of the brightest ($m_{V} = 11.48$ mag), nearest ($z = 0.003326$), and earliest Seyfert galaxies to be discovered \citep{Seyfert1943}, NGC~4151 was one of the best-studied AGNs. This target has been monitored by multiple RM projects in the last three decades including optical RM campaigns \citep{maoz1991, kaspi1996, peterson2004, bentz2006, derosa2018, bentz2022}, and ultraviolet (UV) RM campaigns \citep{Clavel1990, Ulrich1996, Metzroth2006}. Among these projects, \citet{Ulrich1996}, \citet{derosa2018} and \citet{bentz2022} also presented velocity-delay maps of BLR in the UV and optical emission lines, respectively. And these results provide a suitable condition for investigating the kinematic evolution of BLR. Furthermore, NGC~4151 is one of the few AGNs with accretion disk RM \citep{edelson1996, edelson2017}, which limits the size and structure of the disk. Also, it is a changing-look (CL) AGN which usually shows large variability amplitudes in both continuum and emission lines. In the past, NGC~4151 repeated CL phenomenon. For example, in 1984 the spectral type of NGC~4151 changed from type 1 to type 2 \citep{Penston1984}, and \citet{shapovalova2008} found a change from type 1.5 to type 1.8 between 1996 and 2006. Historical data of NGC~4151 show an extreme outburst lasting for more than ten years, and it recently arrived at the second outburst stage. Such unusual variability of continuum should lead to dramatic changes in the radiation pressure on BLR, and then in the properties of geometry and kinematics of BLR, which provide a good opportunity to investigate the nature of BLR during this period. NGC~4151 is only one of two AGNs with several methods for measuring black hole mass (the other is NGC 3227), including gas dynamical modeling \citep{hicks2008}, and stellar dynamical modeling \citep{onken2014, roberts2021}. Thus, NGC~4151 is an ideal candidate to verify the reliability of RM measurement when compared to other results.

In this work, we report the results from the RM campaign of NGC~4151. We successfully measure time lags of broad \ha, \hb, \hg, \hei, and \heii\ emission lines, and obtain their velocity-resolved maps, respectively. In comparison with previous velocity-delay maps, the BLR kinematics of NGC~4151 changed, implying evolutionary kinematics of BLR. Furthermore, we also calculate the black hole mass and dimensionless accretion rate of NGC~4151, and the result indicates a sub-accretor. This paper is arranged as follows. Observations and data reduction are introduced in Section~\ref{sec:obs}. Section \ref{sec:measure} presents light-curve measurements, variability characteristics, and intercalibration of multi-band continuum light curves. Time lag analysis, and measurements of black hole mass and dimensionless accretion rate are given in Section \ref{sec:analysis}. In Section~\ref{sec:discuss}, we discuss the long-term variability trend, kinematics characteristics of BLR, and comparison with previous results in measurements of time lag and black hole mass. Section \ref{sec:conclusion} provides a brief summary.

Throughout the paper, we adopt the distance to NGC~4151 of 15.8 Mpc \citep{yuan2020}.

\section{Observations and Data Reduction} \label{sec:obs}
NGC~4151 was monitored between 2020 November and 2021 June using the 2.4 m telescope of Lijiang Observatory, Yunnan Observatories, Chinese Academy of Sciences. The telescope is located at $100^{\circ}01^{\prime}48^{\pp}$ east longitude and $26^{\circ}41^{\prime}42^{\pp} $ north latitude. This Observatory can be divided into the dry season and rainy season. The rainy season is from July to September with few sunny days, while the rest of the year is the dry season. The weather during the dry season is clear, so we can perform RM observations. Additionally, the telescope is equipped with Yunnan Faint Object Spectrograph and Camera (YFSOC) which can quickly switch observing modes and is convenient. Thus, we carried out photometric and spectroscopic observations each night. The detailed information about the telescope and Observatory were described in \citet{wang2019} and \citet{xin2020}, respectively.

\subsection{Photometry} \label{sec:photo}
\begin{figure}[!ht]
\centerline{
\includegraphics[scale=0.50]{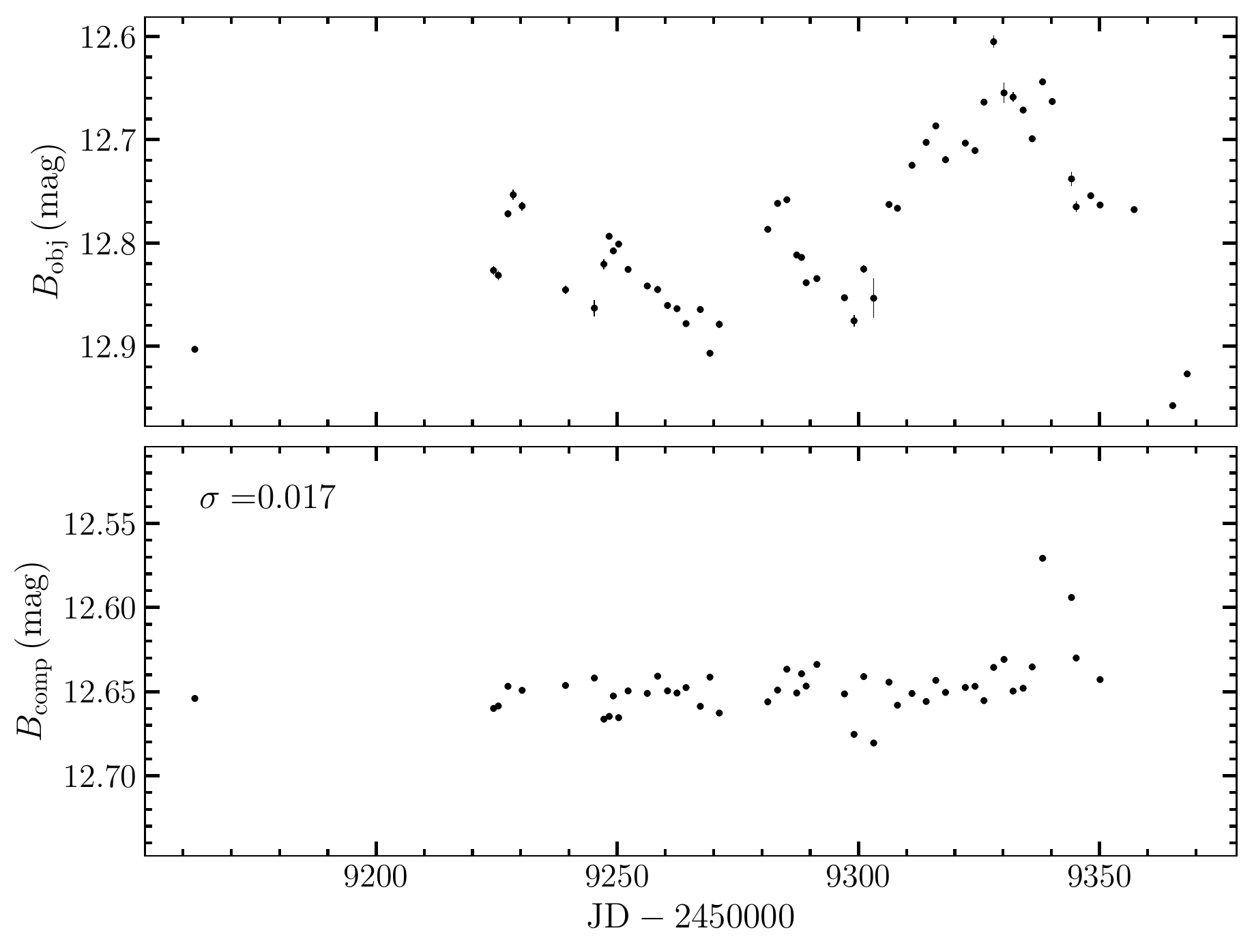}
}
\caption{The photometric light curves of Johnson $B$ filter of NGC~4151 (upper) and comparison star (lower).} 
\label{fig:objcomp}
\end{figure}

Before or after spectroscopic observations, we would use the Johnson $B$ filter and take 30 s exposure to get the broadband image. There are two motivations for performing photometry. Firstly, the photometric data is available for inspecting our spectral data quality. Secondly, we can examine whether the comparison star is unchanging \citep{hu2020a}. The data was reduced by the standard IRAF procedures, including bias subtraction and flat-fielding correction. Since the field of view (FoV) is about $10^{\prime} \times 10^{\prime}$, we could select several stars in the FoV for differential photometry. In addition, we employed a circular aperture with a radius of 4\farcs24 (corresponding to 15 pixels), and the inner and outer radii of the background annulus are 14\farcs15 and 19\farcs81 (corresponding to 50 and 70 pixels), respectively. The size of the aperture was estimated based on the FWHM of stars. Finally, we successfully obtained 55 epochs of photometric observations, and Figure~\ref{fig:objcomp} shows the light curves of our target and comparison star. The standard deviation of the light curve of the comparison star is $\sim$ 0.017 mag, suggesting that it was stable and can be used to calibrate spectral flux.

\subsection{Spectroscopy} \label{sec:spec}
For spectroscopic observations, we applied a slit of 2\farcs5 width to mitigate light loss after taking into account the local average seeing $\sim$ 1\farcs5. The telescope provides various grisms with different resolutions, and we chose Grism 3 possessing a dispersion of 2.86 \AA\,pixel$^{-1}$. Besides that, we also added a UV-blocking filter in observation which can cut off wavelength less than 4150 \AA, accordingly eliminating secondary spectral contamination \citep{feng2021a,feng2021b}. Indeed, taking advantage of the long-slit spectra of the Lijiang 2.4~m telescope, we can simultaneously put NGC~4151 and the comparison star into the slit to adjust the flux of spectroscopy \citep{maoz1990, kaspi2000}. We spent 600 s exposure on each spectrum and utilized IRAF software to process spectra. Furthermore, the extraction aperture and background cover the same region as photometry.

Concerning the flux calibration, we adopted the following steps. First, we picked some spectra of the standard star observed on clear nights to measure the absolute flux of the comparison star. Second, we combined the flux calibrated spectra of the comparison star to generate a template spectrum. Finally, we compared the individual spectrum of the comparison star with the template spectrum to create a response function, which was used for NGC~4151 spectral calibration. As above-mentioned (Section \ref{sec:photo}), the comparison star is invariable and can be employed for calibration. At last, we obtained 41 epochs spectra, one of which was taken using 5\farcs05 width slit. The spectral average signal-to-noise ratio at the rest-frame 5100 \AA\, is $\sim$ 149 pixel$^{-1}$. 

\subsection{Other Telescopes}
Nowadays, with the development of time-domain survey projects, such as the All-Sky Automated Survey for Supernovae (ASAS-SN\footnote{\url{http://www.astronomy.ohio-state.edu/asassn/index.shtml}}) and the Zwicky Transient Facility (ZTF\footnote{\url{https://www.ztf.caltech.edu/}}), it is convenient to increase the cadence and length of observation or to explore the long-term variability. ASAS-SN is used to image the whole sky, and the current depth can reach 18 mag. It consists of multiple stations, and provides 3 dithered 90~s exposures in $V$ or $g$ bands \citep{shappee2014, kocha2017}. Aperture photometry is applied to handle the images with an aperture radius of 16\farcs0.

ZTF is another survey using a 48-inch telescope at Palomar observatory with 47~deg$^{2}$ FoV to monitor transients and variable astronomy phenomenons \citep{Bellm2019, Graham2019, Masci2019}. It scans the Northern Sky on a three-day cadence and the Galactic Plane every night. The depths of $g$ and $r$ bands in 30~s exposures are 20.8 and 20.6 mag, respectively. In addition, the image quality for $g$ and $r$ are 2\farcs1 and 2\farcs0, respectively. But in this work, we adopt the Automatic Learning for the Rapid Classification of Events (ALeRCE\footnote{\url{https://alerce.online}}) API to get accurate ZTF data. ALeRCE has succeeded in managing the ZTF alert stream before it is developed for the Legacy Survey of Space and Time \citep{forster2021, sanchez2021}.

\begin{deluxetable*}{lcccccccccc}[!htbp]
 \tablecolumns{10}
\tablewidth{\textwidth}
\tabletypesize{\scriptsize}
\tablecaption{The Broad Emission-line and Photometric Light Curves}
\label{table:lc}
\tablehead{\multicolumn{6}{c}{Spectra}        &
      \colhead{}                         &
      \multicolumn{3}{c}{Photometry}     \\ 
      \cline{1-6} \cline{8-10}  
      \colhead{JD - 2,450,000}                       &
      \colhead{$F_{\rm H\alpha}$}               &
      \colhead{$F_{\rm H\beta}$}         &
      \colhead{$F_{\rm H\gamma}$}               &
      \colhead{$F_\hei$}               &
      \colhead{$F_\heii$}               &
      \colhead{}                         &
      \colhead{JD - 2,450,000}     &
      \colhead{mag}          &
      \colhead{$\rm Obs$}             
      }
\startdata
9162.44 & $147.136 \pm 1.821$ & $46.207 \pm 0.725$ & $19.227 \pm 0.626$ & $7.533 \pm 0.216$ & $8.290 \pm 1.395$ & & 9153.05 & $12.959 \pm 0.020$ & ZTF \\
9224.38 & $163.628 \pm 1.828$ & $51.484 \pm 0.735$ & $22.135 \pm 0.652$ & $8.387 \pm 0.244$ & $13.503 \pm 1.400$ & & 9156.02 & $12.959 \pm 0.020$ & ZTF \\
9225.37 & $164.933 \pm 1.826$ & $52.237 \pm 0.732$ & $21.685 \pm 0.641$ & $8.742 \pm 0.236$ & $14.133 \pm 1.399$ & & 9158.02 & $12.924 \pm 0.019$ & ZTF \\
9230.31 & $167.195 \pm 1.830$ & $53.336 \pm 0.738$ & $23.489 \pm 0.661$ & $8.966 \pm 0.255$ & $14.731 \pm 1.402$ & & 9162.43 & $12.903 \pm 0.017$ & LJ \\
9239.30 & $169.726 \pm 1.833$ & $52.926 \pm 0.742$ & $22.769 \pm 0.671$ & $9.259 \pm 0.265$ & $13.153 \pm 1.404$ & & 9166.01 & $12.867 \pm 0.019$ & ZTF \\
\enddata
\tablecomments{The emission-line flux is in units of $10^{-13}~\ergscm$. The photometry includes the $B$ band data of Lijiang and the $g$ band data of ZTF, which are intercalibrated by PyCALI, and are also marked with the specific telescope in the ``Obs'' column, ``LJ'' refers to Lijiang 2.4 m telescope, and ``ZTF'' refers to ZTF.
\\
(This table is available in its entirety in a machine-readable form in the online journal.)}
\end{deluxetable*}

\begin{figure*}[!ht]
\centering 
\includegraphics[width=0.49\textwidth]{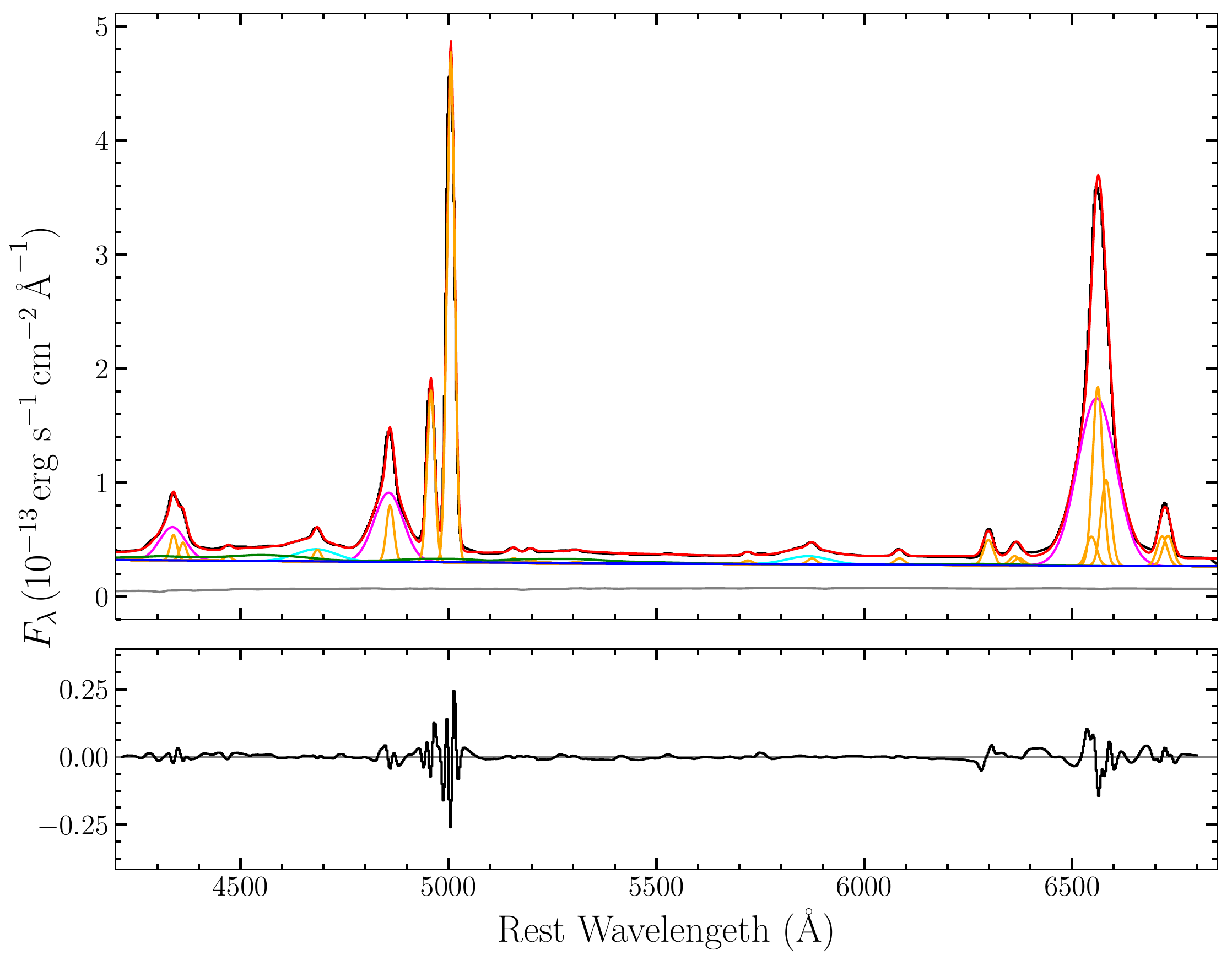}
\includegraphics[width=0.49\textwidth]{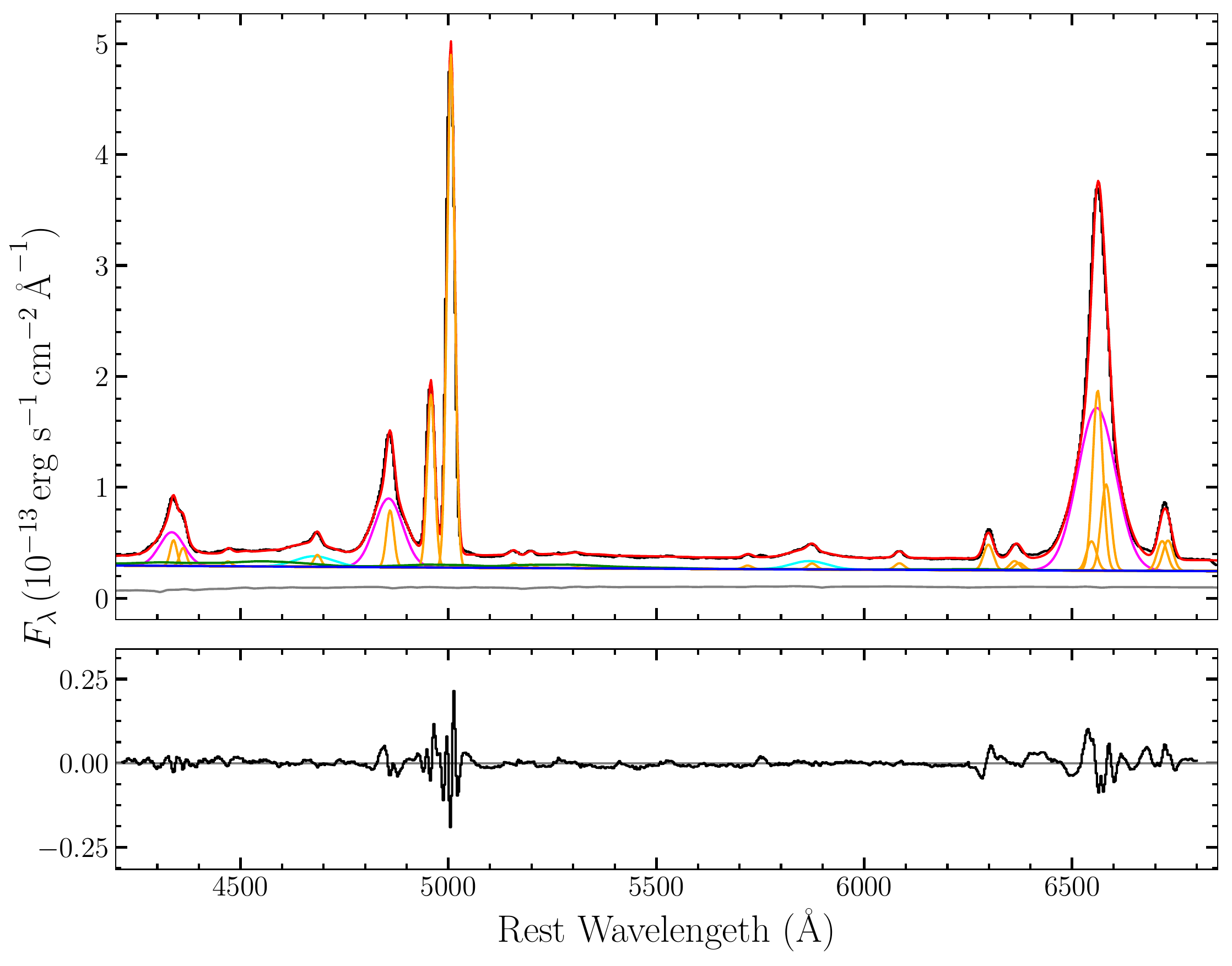}
\caption{The upper left panel shows the fit of the mean spectrum, where the broad Balmer lines (\ha, \hb, \hg) are shown in magenta, the broad \hei\ and \heii\ are shown in cyan, and orange lines represent narrow emission lines. The blue, green, and grey lines correspond to AGN power-law, \feii\ template, and host galaxy template, respectively. Additionally, the red line symbolizes the best-fit model, and the original spectrum is shown in black. The lower left panel is fitting residuals. The right panels are the same as the left panels but for an individual spectrum.} 
\label{fig:fit}
\end{figure*}

\begin{figure*}[!ht]
\centering 
\includegraphics[width=0.9\textwidth]{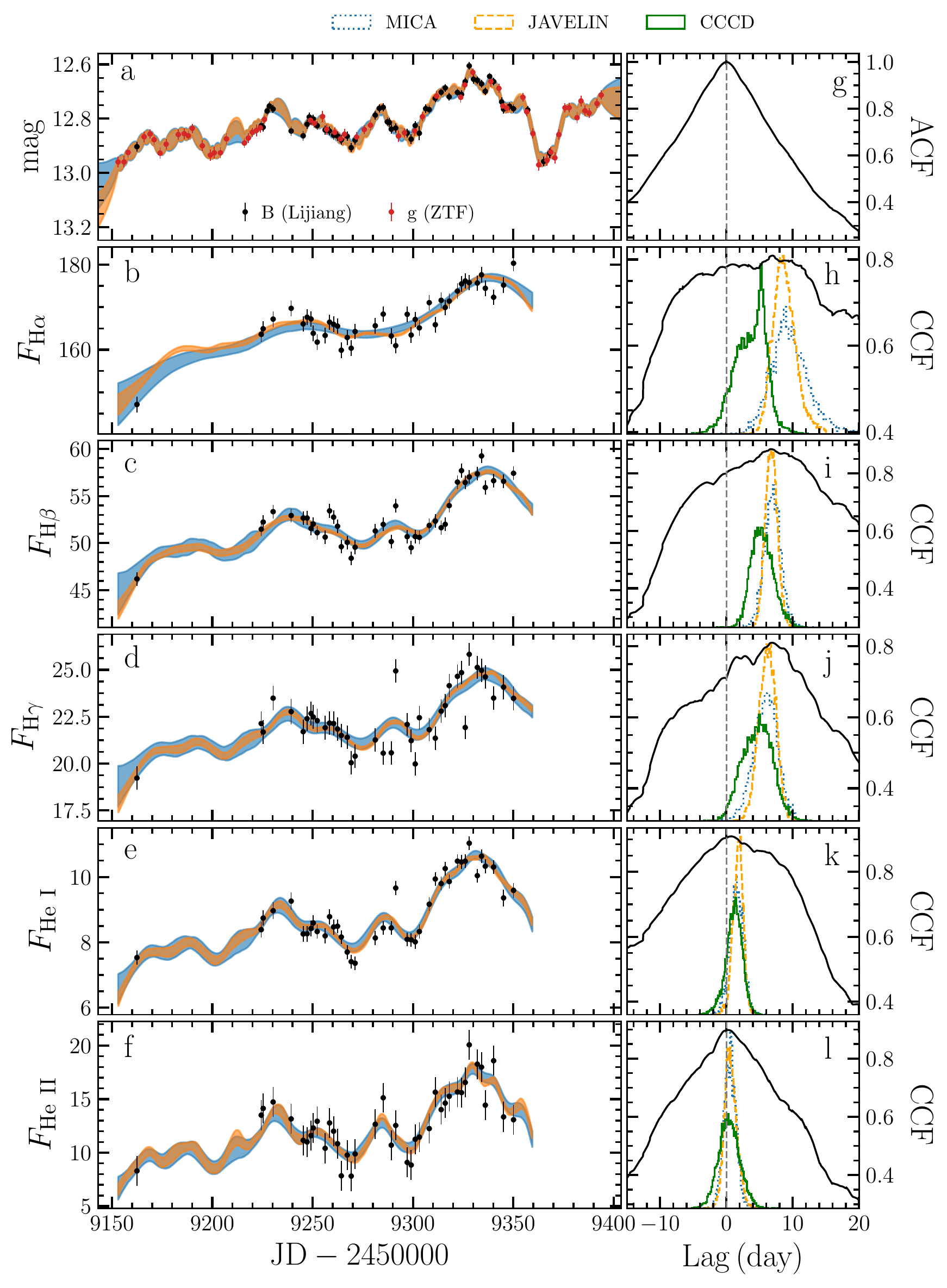}
\caption{The left panels show photometry, \ha, \hb, \hg, \hei, and \heii\ light curves. Panel (a) is the intercalibrated result of $B$ (Lijiang) and $g$ (ZTF). The blue and orange shaded regions represent light curves reconstructed by MICA and JAVELIN. The right panels show time-lag distributions. Panel (g) shows the photometric autocorrelation function (ACF). Furthermore, in panels (h)-(l), the black and green lines correspond to ICCF and CCCD, respectively, and the blue dotted and orange dashed lines represent time-lag probability distributions of MICA and JAVELIN, respectively. The vertical dashed lines represent a time lag of 0 days.
} 
\label{fig:lc}
\end{figure*}

\section{Measurements} \label{sec:measure}
\subsection{Light Curves} \label{sec:lc}
There are two methods to measure light curves of emission lines, one of which is integration and the other is spectral decomposition. For the integration method, we should first select two continuum windows on either side of the emission line and fit them with a straight line as the underlying continuum contribution. Then, the continuum is subtracted from the emission-line profile, and the continuum-subtraction emission-line profile is integrated as emission-line flux. This method is common to extract emission-line flux in RM (e.g., \citealt{kaspi2000, peterson2004, bentz2009, hu2021}), but is not suited for mixed and weak emission lines like \heii\ and \feii. Alternatively, the spectral decomposition can mitigate the contamination of other contributions, because it can decompose pure emission-line components to obtain multi-line light curves (e.g., \citealt{barth2015, hu2015, li2021}). Here, we adopt multi-component spectral fitting to calculate emission-line light curves, and introduce the fitting scheme in the following.

Before the fitting, the galactic extinction and redshift need to be corrected. According to the dust map of \citet{schlafly2011}, the value of $A_{\rm V}$ = 0.074 mag is provided by NED\footnote{\url{http://ned.ipac.caltech.edu}}. We adopted an extinction law of $R_{\rm V}$ = 3.1 to fix reddening \citep{cardelli1989, donnell1994}. Then, each spectrum was shifted to the rest-frame using the \oiii$\lambda$5007. After these corrections, we decomposed the spectra using DASpec\footnote{\url{https://github.com/PuDu-Astro/DASpec}} software with a GUI interface based on the Levenberg-Marquardt technique. We use all spectra except one with slit 5\farcs05 to produce a mean spectrum and fit it at first. The detailed fitting strategy involves several aspects: (1) the host galaxy template from \citet{bc2003}; (2) a \feii\ template from \citet{boroson1992}; (3) a power-law representing AGN continuum; (4) a single-Gaussian function is used to describe each of the broad components of Balmer lines (\ha, \hb\ and \hg) and Helium lines (\hei\ and \heii); (5) moreover, narrow components from the Balmer and Helium lines, and some narrow forbidden lines (see details in \citet{feng2021a}) are depicted by a series of single-Gaussian functions. We compare the widths of \oiii$\lambda$5007 on the mean spectrum and the high-resolution spectrum from HST to estimate the instrumental broadening, which is about 1200~$\kms$. Moreover, the host galaxy and \feii\ in the spectrum are weak and degenerate so that it is not easy to fit the host, therefore we limit the host's FWHM within 1100 and 1300~$\kms$. In the fitting, the widths and shifts of all narrow emission lines are bound to \oiii$\lambda$5007, while the flux ratios of \oiii$\lambda$5007 to \oiii$\lambda$4959 and \nii$\lambda$6583 to \nii$\lambda$6548 are set to 3 and 2.96, respectively. Correspondingly, we first obtain a series of parameters by fitting the mean spectrum, including the spectral index of power law, the FWHM and shift of \feii, and the flux ratios of all narrow lines relative to \oiii$\lambda$5007. And then these parameters are fixed to the fitting of each individual spectrum.
Figure~\ref{fig:fit} shows the decomposition of the mean spectrum (left) and one example of fitting an individual spectrum (right). We also fit the spectrum with a slit width of 5\farcs05, using the same spectral index and flux ratios of narrow lines as the other spectra. 

Finally, the broad emission-line fluxes are measured for the relevant emission lines from the best-fit model to each individual spectrum, and are summarized in Table~\ref{table:lc}. The light-curve error actually contains both Poisson deviation and systematic uncertainty. The systematic errors are caused by the conditions of weather and telescope, which would lead to additional impact on our measurements. We follow \citet{du2014} and apply the median smooth method to assess such errors. The light curves are displayed in the left panel of Figure~\ref{fig:lc}. However, \feii\ is too weak to accurately measure its variability and has not been taken to be analyzed in the present work.

\subsection{Variability Characteristics}
We utilize the following ways to describe the statistical characteristics of the data. The first is the sample mean flux, defined as
\begin{equation}
\langle F \rangle = \frac{1}{N}\sum\limits_{i=0}^N F_{i},
\end{equation}
where $F_{i}$ is $i$-th flux and $\it N$ is the total number of observations. The second is the sample standard deviation, written as
\begin{equation}
S = \bigg(\frac{1}{N-1}\sum\limits_{i=0}^N (F_{i} - \langle F \rangle)^2\bigg)^{1/2}.
\end{equation}
The third is the fractional variability amplitude $F_{\rm var}$ to describe AGN intrinsic variability \citep{Rodr1997}, 
\begin{equation}
F_{\rm var}=\frac{(S^2-\triangle^2)^{1/2}}{\langle F \rangle},
\end{equation}
where $\triangle^2 = \frac{1}{N}\sum\limits_{i=0}^N \triangle_{i}^2$, and $\triangle_{i}$ is the uncertainty of $F_{i}$. According to \citet{edelson2002}, the error $\sigma_{\rm var}$ of $F_{\rm var}$ is expressed as
\begin{equation}
\sigma_{\rm var}=\frac{1}{F_{\rm var}}\bigg(\frac{1}{2N}\bigg)^{1/2}\frac{S^2}{{\langle F \rangle}^2}.
\end{equation}
The last is $R_{\max}$, the ratio between the maximum and the minimum of the light curve. The relevant measurements are listed in Table~\ref{table:stas}. 

\begin{deluxetable}{lcccccccccc}[!ht]
 \tablecolumns{10}
\tablewidth{\textwidth}
\tabletypesize{\scriptsize}
	\tablecaption{Light Curve Statistics \label{table:stas}}
	%\tablecolumns{5}
	\tablewidth{\textwidth}
		\tablehead{
			\colhead{Light Curve} &
			\colhead{Mean Flux} &
			\colhead{Standard Deviation} &
			\colhead{$F_{\rm var}(\%)$} &
			\colhead{$R_{\rm max}$}
		}
\startdata
\ha & 167.73 & 6.06 & $3.49\pm0.42$ & 1.23\\
\hb & 52.78 & 2.86 & $5.30\pm0.62$ & 1.28\\
\hg & 22.48 & 1.55 & $6.39\pm0.84$ & 1.34\\
\hei & 8.98 & 1.00 & $10.95\pm1.27$ & 1.50\\
\heii & 12.95 & 2.92 & $20.12\pm2.86$ & 2.57\\
\enddata
\tablecomments{The emission-line fluxes are in units of $10^{-13}~\ergscm$.}
\end{deluxetable}

\subsection{Intercalibration}  \label{sec:intercali}
We only add the $g$ band of ZTF into the $B$ band of Lijiang as the continuum light curve to analyze time lag, considering that the $r$ band data of ZTF in our monitoring period is few. Before merging the data, we confirm that the time lag between the $g$ and $B$ bands is less than one day. Owing to utilizing different apertures and different filters, the datasets need to be intercalibrated. 
Here, we calibrate multi-band data via PyCALI\footnote{\url{https://github.com/LiyrAstroph/PyCALI}}, which fits light curve through the damped random walk process and uses multiplicative and additive factors to regulate these data into a common scale \citep{li2014L}. And the intercalibrated parameters are obtained based on the Bayesian framework of the diffusive nested sampling algorithm \citep{Brewer2011}. Finally, the calibrated data are presented in the panel (a) of Figure~\ref{fig:lc} as the continuum light curve used in subsequent analyses, and the data are given in Table~\ref{table:lc}.

\begin{figure*}[!ht]
\centering 
\includegraphics[width=1\textwidth]{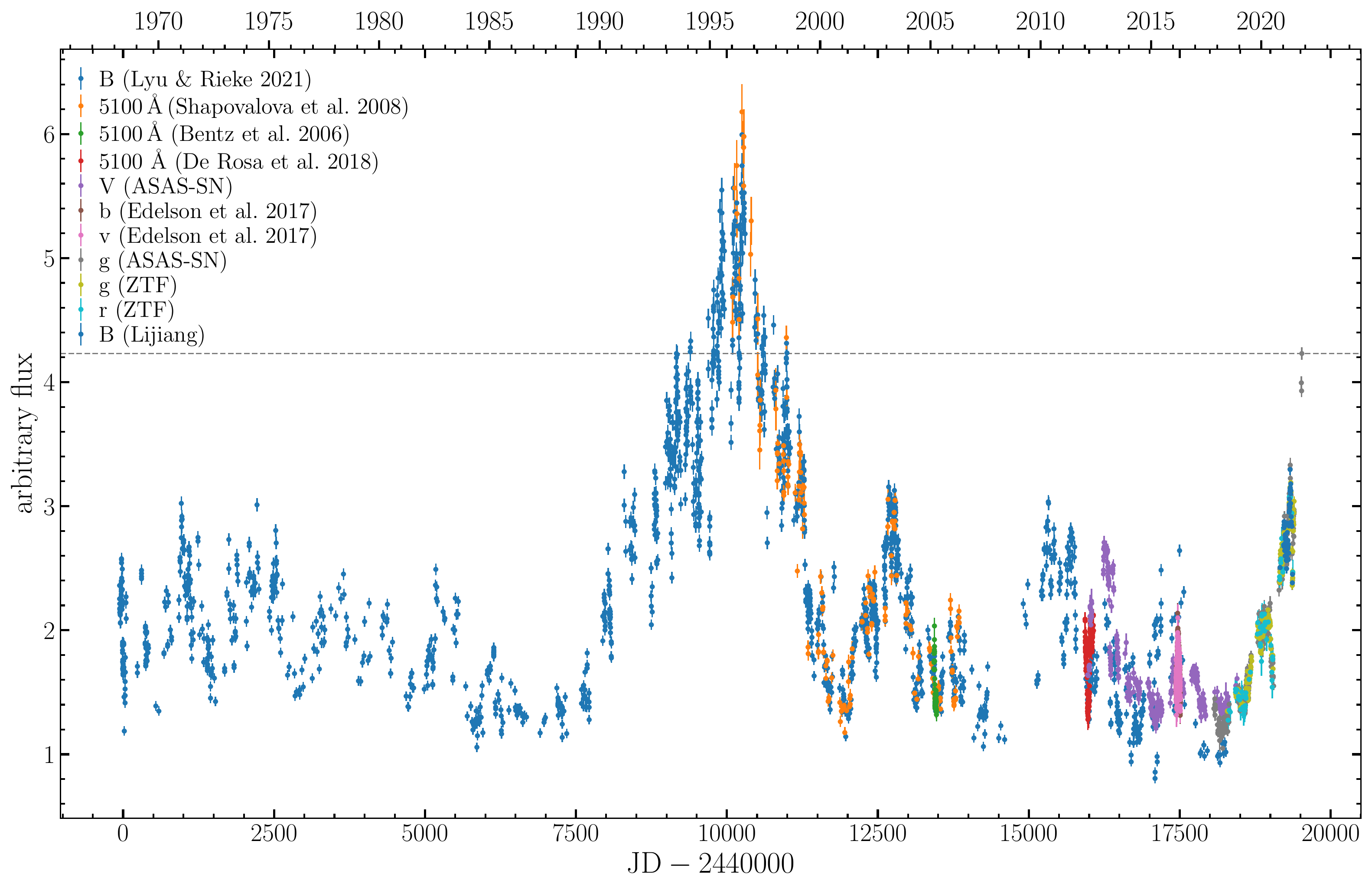}
\caption{This is the long-term light curve of NGC~4151, which was generated by combing the collected continuum and photometry with the $B$ band of Lijiang. The details about the intercalibrated data can be seen in Section~\ref{sec:intercali}. The horizontal dashed line represents the latest peak.} 
\label{fig:calilc}
\end{figure*}

In addition, we collect the historical light curves of NGC~4151 from previous literature to probe the long-term variations of NGC~4151, involving 5100 \AA\ continuum \citep{bentz2006, shapovalova2008, derosa2018}, $B$ band \citep{lyu2021}, $b$ and $v$ bands \citep{edelson2017}, as well as sky survey data, like $V$ and $g$ bands of ASAS-SN, $g$ and $r$ bands of ZTF. Similarly, we also scale and shift these data with the $B$ band of Lijiang through PyCALI (see Figure~\ref{fig:calilc}). 

\begin{deluxetable}{lcccccccccc}[!ht]
\tablewidth{\textwidth}
\tabletypesize{\scriptsize}
\tablecaption{Rest-frame Time Lags \label{table:lag}}
\tablewidth{\textwidth}
\tablehead{
    \colhead{$\rm Line$} &
	\colhead{$r_{\rm max}$} &
	\colhead{$\tau_{\rm cent}$} &
	\colhead{$\tau_{\rm MICA}$} &
	\colhead{$\tau_{\rm JAV}$}&
	\colhead{$\tau_{\rm Mean}$}
		}
\startdata
\ha & 0.81 &  $5.00_{-3.80}^{+0.84}$ & $9.28_{-2.50}^{+2.91}$ &  $8.60_{-1.55}^{+1.80}$ &  $7.63_{-2.62}^{+1.85}$ \\ 
\hb &0.88 &  $4.95_{-1.20}^{+2.12}$ & $6.94_{-1.27}^{+1.21}$ &  $6.73_{-0.91}^{+0.91}$ &  $6.21_{-1.13}^{+1.41}$ \\ 
\hg & 0.81 &  $4.90_{-2.61}^{+1.99}$ & $5.95_{-1.88}^{+1.69}$,  &  $6.16_{-1.33}^{+1.26}$ &  $5.67_{-1.94}^{+1.65}$ \\ 
\hei & 0.91 &  $1.46_{-1.52}^{+0.86}$ & $1.52_{-1.00}^{+1.02}$ &  $1.78_{-0.83}^{+0.70}$ &  $1.59_{-1.11}^{+0.86}$ \\ 
\heii & 0.90 &  $0.23_{-1.39}^{+1.76}$ & $0.55_{-0.94}^{+0.91}$ &  $0.59_{-0.84}^{+1.00}$ &  $0.46_{-1.06}^{+1.22}$ \\
\enddata
\tablecomments{$\tau_{\rm Mean}$ is the mean value of $\tau_{\rm cent}$, $\tau_{\rm MICA}$, and $\tau_{\rm JAV}$ in units of days.}
\end{deluxetable}

\begin{figure*}[!ht]
\centering 
\includegraphics[width=0.33\textwidth]{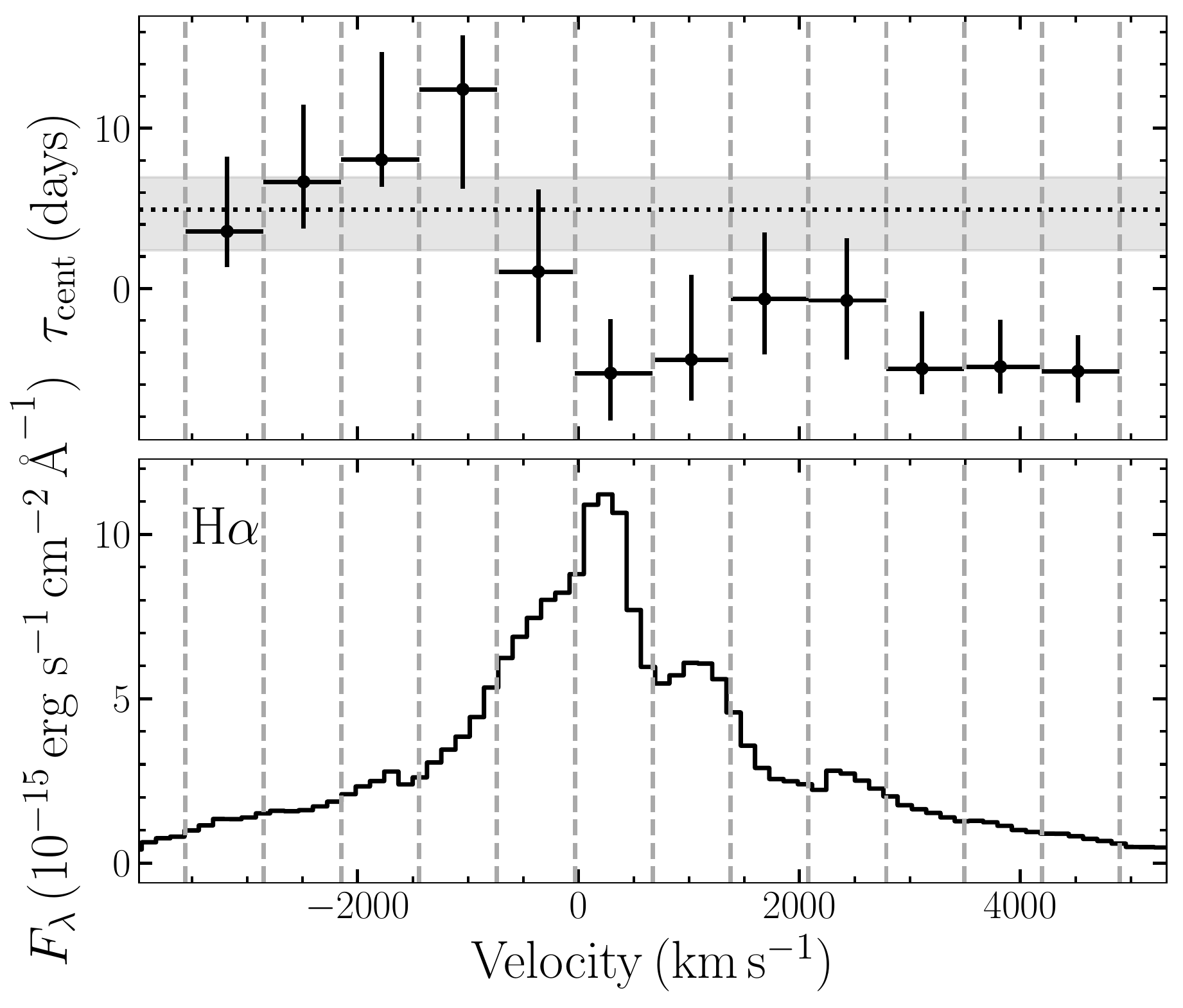}
\includegraphics[width=0.33\textwidth]{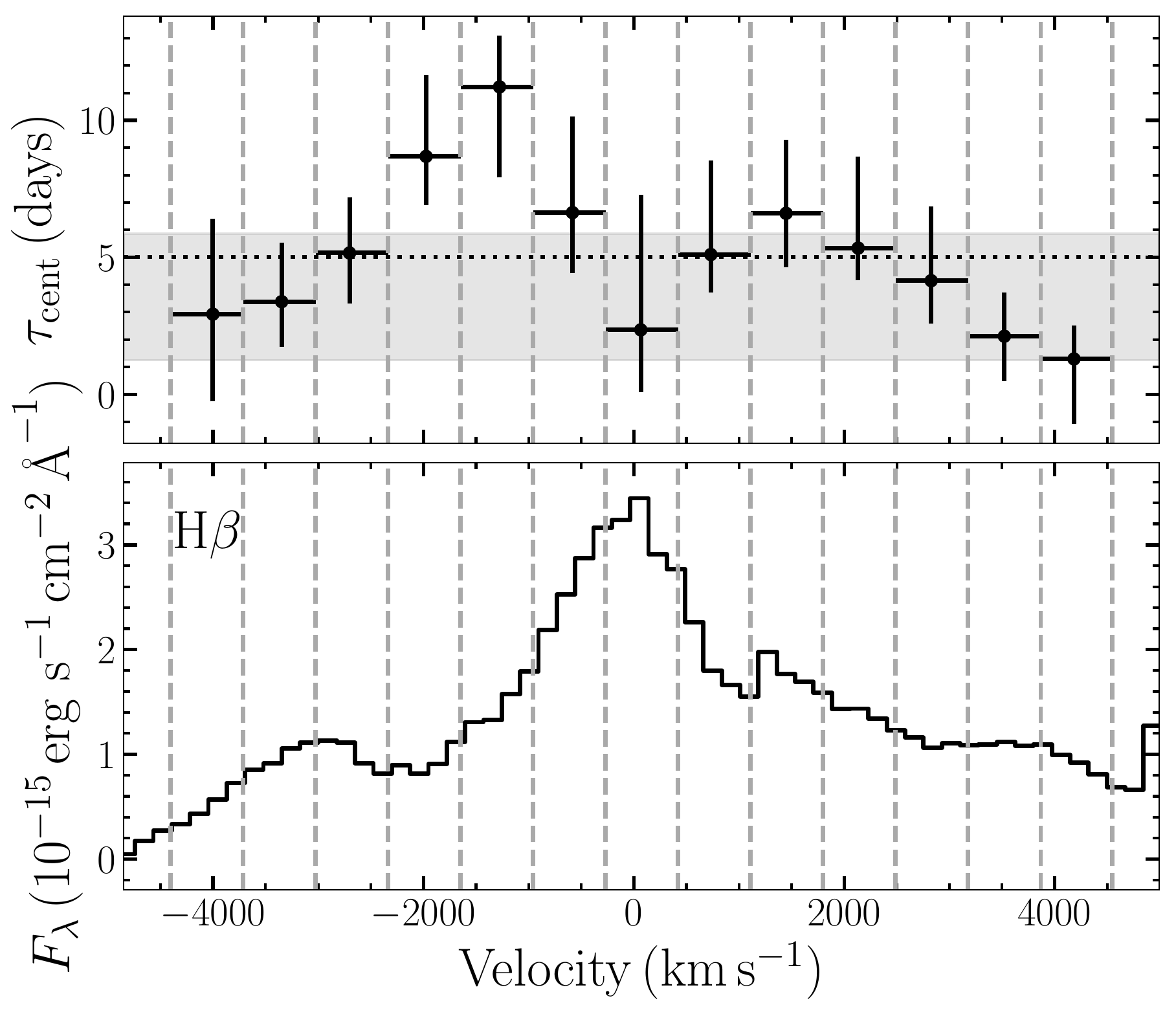}
\includegraphics[width=0.33\textwidth]{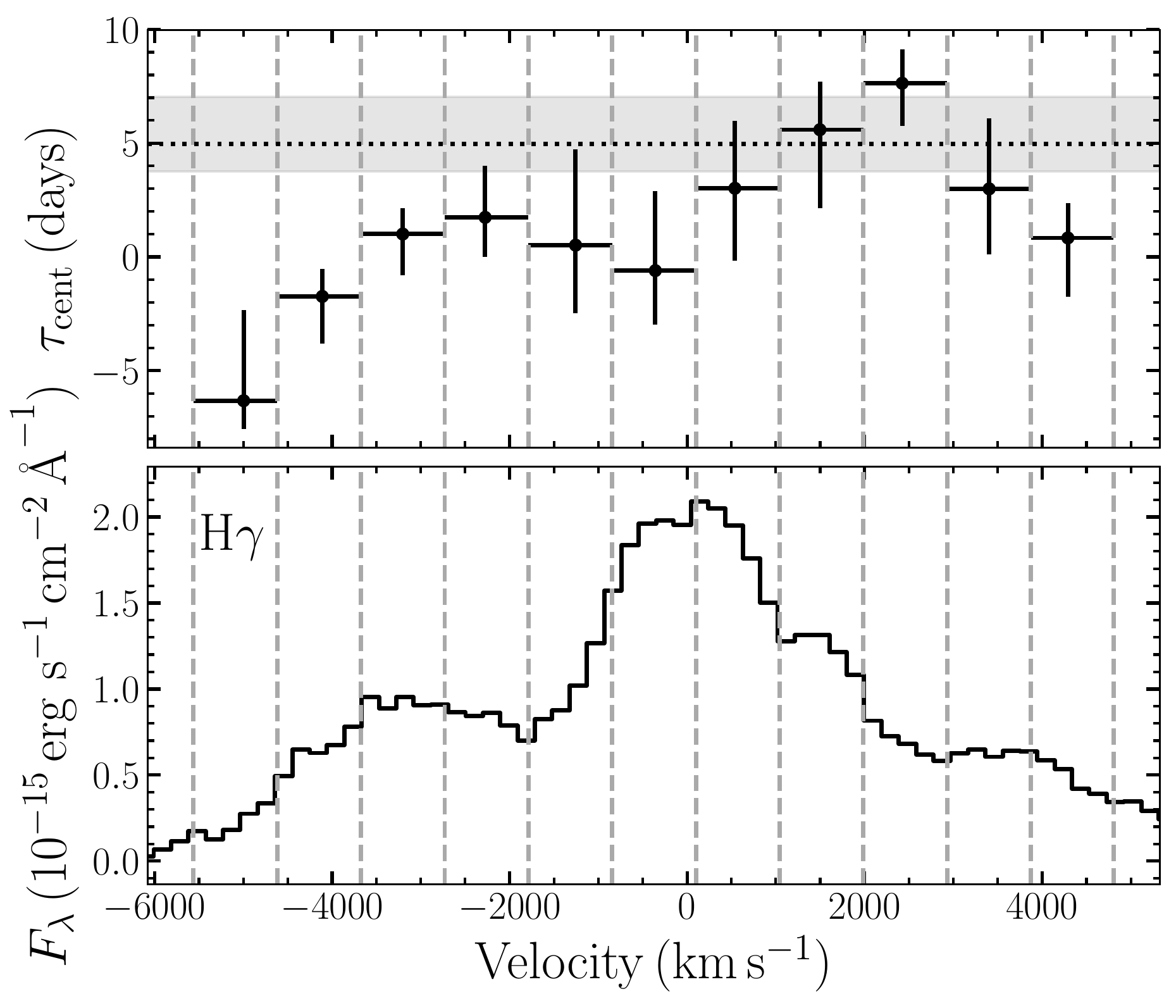}
\includegraphics[width=0.33\textwidth]{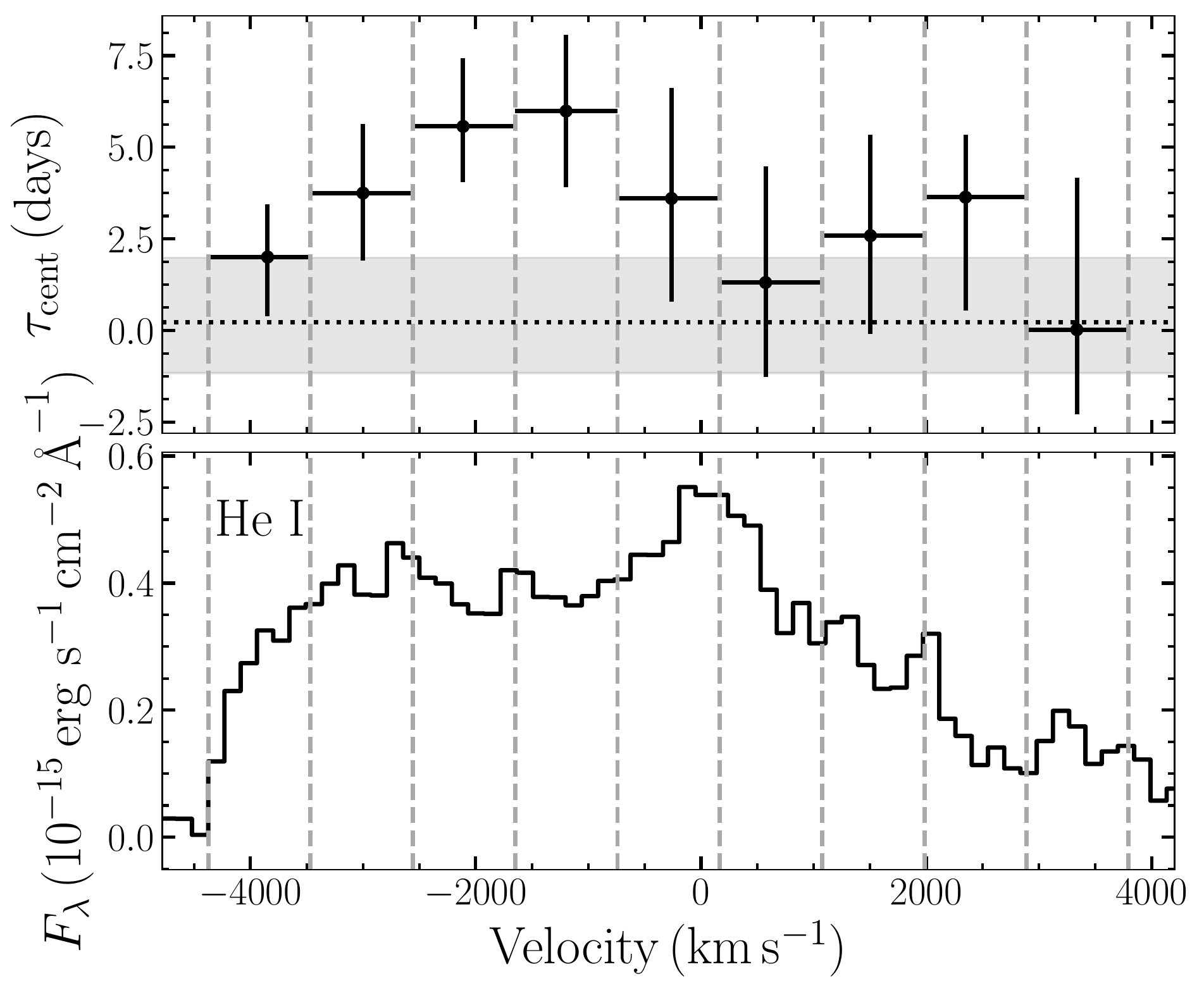}
\includegraphics[width=0.33\textwidth]{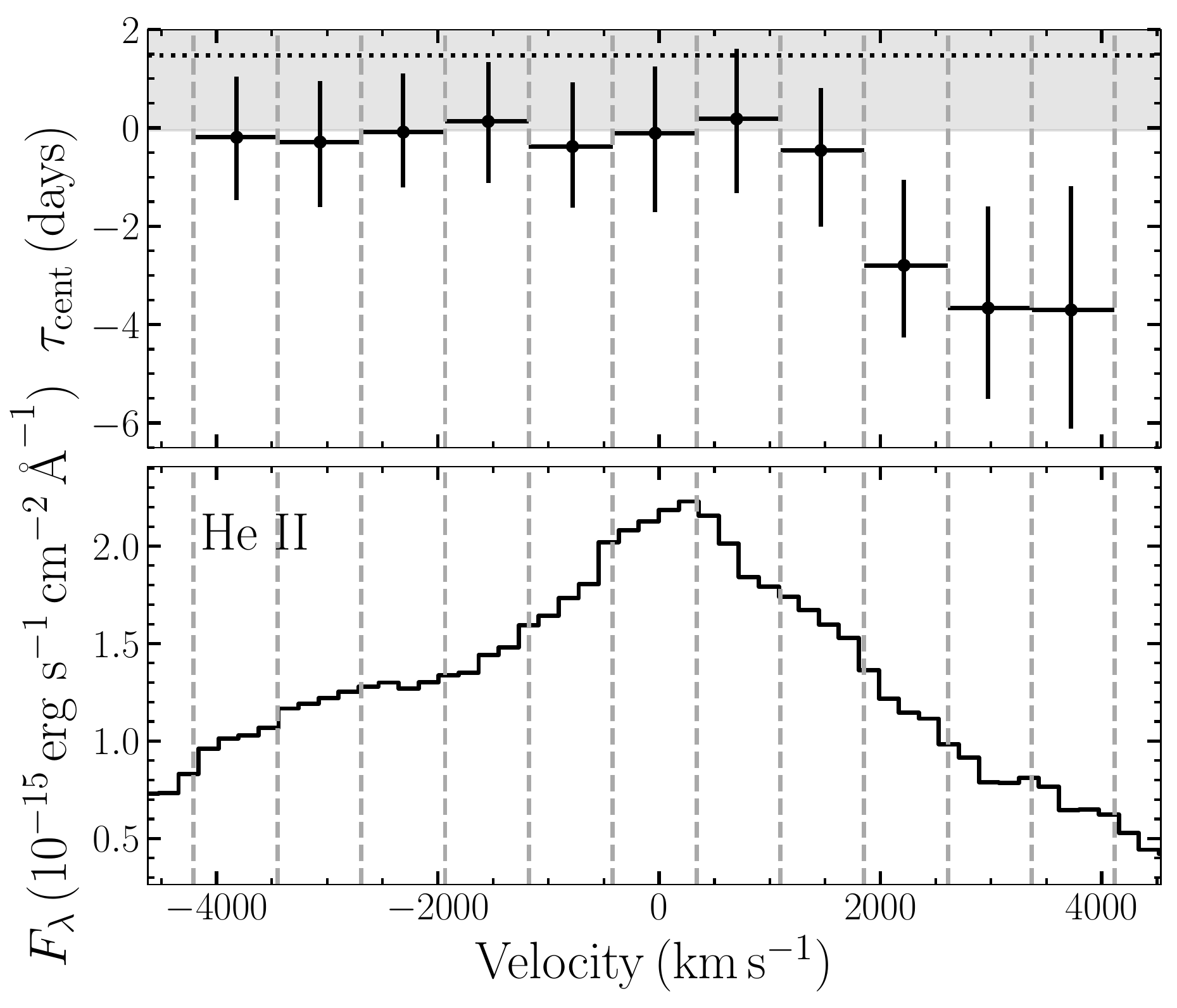}
\caption{The velocity-resolved time-delay results of \ha, \hb, \hg, \hei, and \heii. The top panel in each subgraph shows the velocity-resolved structure for each emission line. In each panel, the horizontal dotted line and grey band represent the ICCF measured lag and CCCD estimated errors, respectively. The vertical dashed lines denote the edges of each velocity bin. The bottom panel in each subgraph shows the rms spectrum of each emission line, for which the continuum is subtracted.}
\label{fig:velocitylag}
\end{figure*}

\section{Analysis and Results} \label{sec:analysis}
\subsection{Mean Time Lags} \label{sec:lag}
In order to investigate the responses of broad emission lines to continuum variations, we take the intercalibrated photometry light curve in Section~\ref{sec:intercali} as the continuum light curve used to analyze time lags of emission lines with respect to the continuum variations. Three approaches are adopted to estimate time lags between emission lines and continuum, which include the interpolated cross-correlation function \citep[ICCF;][]{Gaskell1986,Gaskell1987}, MICA\footnote{\url{https://github.com/LiyrAstroph/MICA2}} \citep{li2016}, and JAVELIN \citep{zu2011}. 

Traditionally, the ICCF is the most commonly used method for measuring time lags, and we choose the centroid of a region with the correction coefficients $r \geq 0.8 r_{\rm max}$ as a time lag $\tau_{\rm cent}$, where $r_{\rm max}$ is the maximum correlation coefficient. Similar to \citet{peterson1998, peterson2004}, we utilize the ``flux randomization'' (FR) and ``random subset selection'' (RSS) procedures to determine the lag uncertainties. Accordingly, a cross-correlation centroid distribution (CCCD) is built by $10^4$ Monte Carlo realizations of the FR/RSS procedures, and we set the 15.87\% and 84.13\% quantiles of the CCCD as the lower and upper bounds of the lag.
 
MICA and JAVELIN are dependent on the damped random walk model, while the specific shapes of transfer functions are different. Specially, MICA and JAVELIN adopt the superposition of a set of moved Gaussian functions and a top-hat function as transfer function, respectively. For simplicity, we apply one Gaussian as the transfer function of MICA. The Gaussian and the top-hat centers stand for time delays in MICA and JAVELIN, respectively. Additionally, their time-lag posterior probability distributions are evaluated by the Monte Carlo Markov Chain (MCMC) technique. The medians of their distributions are taken as time delays for MICA and JAVELIN, respectively. The lower and upper uncertainties of the time delays are determined by the 15.87\%, and 84.13\% quantiles of the relevant distributions. Besides that, both MICA and JAVELIN can reconstruct light curves, and the reconstructed results are filled with blue and orange blocks individually, as shown in the left panels of Figure~\ref{fig:lc}.

The time lags calculated from these three ways are listed in Table~\ref{table:lag}, where $\tau_{\rm MICA}$ and $\tau_{\rm JAV}$ are the time lags measured by MICA and JAVELIN, respectively. Additionally, the time-lag posterior probability distributions are demonstrated in the right panels of Figure~\ref{fig:lc}. Remarkably, the results of MICA and JAVELIN are consistent with each other, and are also in agreement with the ICCF measurements except \ha\ and \hb. Overall, the time lags measured by the ICCF method are lower than the results of the other two methods. This may be that both MICA and JAVELIN describe the light curve based on the damped random walk model, while ICCF directly linearly interpolates the light curve. Therefore, the measured lags will be affected by the principle difference between the ICCF and the MICA and JAVELIN, as well as some scattered points or gaps on the light curve, which may be why the ICCF lags are generally lower than the MICA and JAVELIN lags. We also find that time lags measured by each method follow $\tau_{\rm H\alpha} > \tau_{\rm H\beta} > \tau_{\rm H\gamma} > \tau_{\rm He\ I} >\tau_{\rm He\ II}$, which are radial stratification related to optical depths \citep{Korista2004} and ionization energy \citep{CS1988}. We adopt the mean values of their respective three time lags in the subsequent analyses. We also compute the ratios of the mean lags of \ha, \hb, \hg, \hei, and \heii\ relative to \hb\ to be $1.23 : 1.00 : 0.91 : 0.26 : 0.07$.

\subsection{Velocity-resolved Lags}
The time lags between continuum and broad emission lines, measured in Section~\ref{sec:lag}, characterize the mean sizes of the BLRs. By analyzing time series at different line-of-sight velocity bins, that is, velocity-resolved time lags, it is possible to yield some information about the kinematics and geometry of BLR. Note that the results of this approach are not necessarily accurate, as weak reverberation effects are ignored using the mean time lag for each bin \citep[see][]{derosa2018}. There are three patterns to depict the kinematics of BLR, including virialized, inflowing, and outflowing motions. If lags at the line center are longer than those at line wings, the BLR is a virialized motion. When the lags of blueshifted velocity bins are longer than those of redshifted velocity bins, the BLR is an infall model, otherwise, it is the signature of outflow.   

Below are the detailed steps for velocity-resolved RM. First, we select two windows on both sides of the emission line on the rms spectrum to fit the underlying continuum with a straight line. Second, we divide each continuum-subtracted emission line into several velocity-space bins with an equal velocity interval. Then we apply these bins in individual spectrum to obtain their light curves, which are shown in the Appendix~\ref{sec:appendix}. Finally, we perform ICCF between the light curves of emission line and continuum for each bin to measure lag with uncertainties estimated by CCCD. The velocity-dependent lags are shown in Figure~\ref{fig:velocitylag}. Indeed, we employ a different method from Section~\ref{sec:lc} to decompose emission lines, and this method used here can mitigate the degeneracy in the line wings introduced by the multi-component spectral fitting. Furthermore, the measurements from each method are generally consistent.

We note that velocity-delay structures of \ha, \hb, \hg, and \hei\ are complicated. All the four emission lines show the double-peaked profiles of velocity-delay structures, but are not identical to each other. The blue-side peaks are larger than the red-side ones in \ha, \hb, and \hei, but \hg\ appears to have a larger red-side peak. For \hg, it may be affected by \oiii\,$\lambda$4363. Additionally, \heii\ shows an infalling motion, however it seems unreliable. In the individual spectrum, \heii\ is weak and blended with \feii, making it hard to acquire a reliable velocity-delay structure. In short, since the velocity-resolved lags of \ha, \hb, and \hei\ are consistent, we treat this kind of signature as the BLR kinematics, which may be a combined effect between virial and inflow motions. Mrk~6 also shows a similar velocity-delay structure \citep{doroshenko2012, grier2013b, du2018}. However, in 2012 \citep{derosa2018}, the pattern of NGC~4151 is virialized motion, perhaps indicating a change in the kinematics of BLR. In addition, \citet{bentz2022} reanalyzed RM data from NGC~4151 in 2005, and modeled these data to constrain the geometry and kinematics of BLR. The results indicate that the BLR can be characterized by a thick disk with opening angle $\sim 57^{\circ}$ and inclination angle $\sim 58^{\circ}$, and the kinematics of BLR are dominated by eccentric bound orbits, about 10\% of the orbits tending to be near-circular motions. The above results further support that the kinematics of BLR have changed, as well as the geometry of BLR.

\begin{deluxetable*}{lcccccccccc}[!ht]
\tablewidth{\textwidth}
\tabletypesize{\scriptsize}
\tablecaption{Time Lags, Line Widths, and Black Hole Masses \label{table:width}}
\tablewidth{\textwidth}
\tablehead{
	\multirow{2}{*}{$\rm Line$} &
    \multicolumn{1}{c}{$\tau_{\rm Mean}$} &
     \multicolumn{1}{c}{$\tau_{\rm Tot}$} &
	\multicolumn{1}{c}{FWHM} &
	\multicolumn{1}{c}{$\sigma_{\rm line}$} &
	\multicolumn{1}{c}{$M_{\rm vir}$}&
	\multicolumn{1}{c}{$M_{\bullet}$}&
	\multicolumn{1}{c}{$M_{\bullet}^{'}$}\\
	\colhead{}&
	\multicolumn{2}{c}{(days)} &
	\multicolumn{2}{c}{($\kms$)}&
	\multicolumn{3}{c}{($\times 10^7 M_{\odot}$)}
		}
\startdata
\ha & $7.63_{-2.62}^{+1.85}$ & $8.45_{-2.96}^{+2.17}$ & $4897\pm 5$ & $ 2069\pm 2$ & $3.57_{-1.23}^{+0.87}$ &$ 4.64_{-1.59}^{+1.13}$ &$ 5.14_{-1.80}^{+1.32}$\\ 
\hb & $6.21_{-1.13}^{+1.41}$ & $7.03_{-1.47}^{+1.73}$ & $5003\pm 13$ & $ 2074\pm 5$ & $3.03_{-0.55}^{+0.69}$ &$ 3.94_{-0.72}^{+0.90}$ &$ 4.47_{-0.93}^{+1.10}$\\ 
\hg & $5.67_{-1.94}^{+1.64}$ & $6.50_{-2.28}^{+1.96}$ & $4799\pm 20$ & $ 2011\pm 8$ & $2.55_{-0.87}^{+0.74}$ &$ 3.31_{-1.13}^{+0.96}$ &$ 3.80_{-1.33}^{+1.15}$\\ 
\hei & $1.59_{-1.11}^{+0.86}$ & $2.41_{-1.45}^{+1.18}$ & $6031\pm 50$ & $ 2340\pm 11$ & $1.13_{-0.79}^{+0.61}$ &$ 1.46_{-1.03}^{+0.79}$ &$ 2.23_{-1.34}^{+1.09}$\\ 
\heii & $0.46_{-1.06}^{+1.22}$ & $1.28_{-1.40}^{+1.54}$ & $7099\pm 52$ & $ 2375\pm 6$ & $0.45_{-1.04}^{+1.20}$ &$ 0.58_{-1.35}^{+1.56}$ &$ 1.64_{-1.78}^{+1.97}$\\ 
\enddata
\tablecomments{$\tau_{\rm Mean}$ is the mean lag of three methods in Section~\ref{sec:lag}, $\tau_{\rm Tot}$ is the result of $\tau_{\rm Mean}$ plus $\tau_{\rm b-uv}$ (see details in Section~\ref{sec:uv/optical}). FWHM and $\sigma_{\rm line}$ represent line widths of mean spectrum. $M_{\rm vir}$ is the virial product, $M_{\bullet}$ is the black hole mass when adopting $f =1.3$ and not considering the error of $f$, and $M_{\bullet}^{'}$ is measurement with $\tau_{\rm Tot}$.}
\end{deluxetable*}

\subsection{Black Hole Mass and Accretion Rate} \label{sec:bhmass}  
Assuming that BLR clouds are bound by the gravity of the black hole and follow the virialized motion, the black hole mass can be expressed as  
\begin{equation}
M_{\bullet} = f \frac{R_{\rm BLR} (\Delta V)^2}{G} \equiv f M_{\rm vir}, 
\end{equation}
where $R_{\rm BLR} = c \tau$ represents the mean size of BLR, $c$ is the speed of light, $\tau$ is the mean time lag of the BLR, $G$ is the gravitational constant, $\Delta V$ is the broad emission-line velocity, usually characterized by the emission-line width, $f$ is the virial factor, depending on geometry, kinematics, and inclination of the BLR, and $M_{\rm vir}$ is virial product.

The line width is generally depicted by FWHM or line dispersion $\sigma_{\rm line}$ of a mean or rms spectrum. $\sigma_{\rm line}$ is expressed as 

\begin{equation}
 \sigma_{\rm line}^{2}(\lambda) =  \frac{\int \lambda^{2} F(\lambda)d\lambda}{\int F(\lambda)d\lambda}  - \bigg\lbrack \frac{\int \lambda F(\lambda)d\lambda}{\int F(\lambda)d\lambda} \bigg\rbrack^{2},
\end{equation}
where $F(\lambda)$ is the flux density of emission-line profile at $\lambda$. Here, we present line-width measurements of a mean spectrum, and utilize the FWHM to estimate $M_{\bullet}$. Because the rms spectrum contains narrow line residuals and produces unreliable line widths, we do not use them here, but instead introduce them in Section~\ref{sec:vr}.

We use the bootstrap technique to assess the line width and corresponding uncertainties. Specifically, we randomly select $\it N$ spectra with replacement from $\it N$ spectra, and then create a new mean spectrum after removing the repeated spectra. This process is repeated 1000 times to generate 1000 new mean spectra. Regarding each new mean spectrum, we adopt the spectral decomposition method and employ the best-fitting model in Section~\ref{sec:lc} to fit it, thereby acquiring FWHMs of broad emission lines from the fitted Gaussian models. The FWHMs and errors of emission lines are calculated by the median and the standard deviation of these measurements. Then, we subtract the fitting models except for broad emission-line components from the aforementioned mean spectra, and this subtraction generates the corresponding residual spectra. The distribution of $\sigma_{\rm line}$ is measured from the residual spectra for each broad emission line. The median and the standard deviation of this distribution are regarded as the measured values of $\sigma_{\rm line}$ and its error, respectively. The above FWHM and $\sigma_{\rm line}$ also need to subtract instrumental broadening contribution in quadrature, which is about 1200 $\kms$ measured in Section~\ref{sec:lc}, and the corrected line widths are summarized in Table~\ref{table:width}. 

The $f$ factor is estimated from the following methods: the black hole mass and stellar velocity dispersion ($M_{\bullet}-\sigma_{*}$) relation \citep[e.g.,][]{onken2004, woo2010, grier2013a, ho2014}, the BLR dynamical modeling analysis in a single AGN \citep{pancoast2012, li2013, li2018}, the widths and shifts of redward shifted broad emission lines \citep{liu2017,liu2022,MJ20}, and fitting the spectral energy distributions with the accretion disk model \citep{mej2018}. \citet{Kormendy2013} showed that $M_{\bullet}-\sigma_{*}$ relation is related to the bulge type, thus we adopt the $f$ factor obtained from \citet{ho2014}, who derived it by subdivided bulge types. Due to a classical bulge in NGC~4151 \citep{ho2014}, we utilize $f = 1.3$, $\tau_{\rm mean}$, and FWHM in the mean spectrum to measure $M_{\bullet}$, and also provide $M_{\rm vir}$ in Table~\ref{table:width}. We note that the black hole masses measured by the different emission lines are consistent with each other within errors, though the masses estimated from the Balmer lines are larger than those from \hei\ and \heii. 

The dimensionless accretion rate is defined as $\dotm=\dot{M}_{\bullet}\,c^2/L_{\rm Edd}$, where $\dot{M}_{\bullet}$ is the mass accretion rate and $L_{\rm Edd}$ is the Eddington luminosity \citep{wang2014}. Based on the standard accretion disk model \citep{Shakura1973}, $\dotm$ can be estimated by the following formula \citep{wang2014}
\begin{equation}\label{equ:accre}
\dotm = 20.1\,\left(\frac{\ell_{44}}{\cos i}\right)^{3/2}m_7^{-2},
\end{equation}
where $\ell_{44}=L_{5100}/10^{44} \ergs$, $m_7 = M_{\bullet}/10^7\sunm$, and $i$ is the inclination of the accretion disk. Here, we adopt $i=45 \pm 5^{\circ}$ that is the inclination of narrow-line region (NLR) in NGC 4151 \citep{Das2005}, which agrees with the inclination of the NLR bicone model reported by \citet{Fischer2013}. Note that the inclination of NLR does not always coincide with accretion disk, BLR or torus. For instance, the inclination $i=45^{\circ}$ disagrees with an average inclination of the accretion disk to be $\sim 20^{\circ}$ derived from Fe K$\alpha$ modeling \citep{Nandra1997}. From the fitting results of all individual spectra, we extract the AGN continuum component and obtain a mean flux of AGN continuum at 5100 \AA, $29.95 (\pm 5.99) \times 10^{-15} \ergscma$, and then a luminosity of $L_{5100} = 4.56 (\pm 0.91) \times 10^{42} \ergs$. Combining this luminosity with $M_{\bullet}=3.94_{-0.72}^{+0.90} \times 10^7 \sunm$ measured from \hb, the value of $\dotm$ is $0.02_{-0.01}^{+0.01}$ (not considering uncertainties of inclination), implying that NGC~4151 is a sub-Eddington accretor.

\section{Discussion} \label{sec:discuss}
\subsection{Long-term Variability Trend} \label{sec:long-term}
We collect the historical light curves of NGC~4151 with a temporal baseline of $\sim$53 years, and intercalibrate them in Section~\ref{sec:intercali}, as displayed in Figure~\ref{fig:calilc}. The light curve consists of a major outburst and a series of minor flares. And, some of these peaks and valleys are accompanied by different spectral types, i.e., multiple CL phenomena. For example, the spectral type of NGC~4151 is an intermediate type in \citet{Osterbrock1976}, but is close to type 2 in 1984 while the flux is also fading to a minimum \citep{Penston1984}. When the flux reaches the maximum in 1996, the spectral type is Seyfert 1.5. The type changes to Seyfert 1.8 at the two minimum states in 2001 and 2005 \citep{shapovalova2008}. Interestingly, during the period of $1996-1999$, NGC~4151 has strong continuum, but the emission line flux is saturated and does not respond to continuum variations \citep{shapovalova2008}. 

Previous optical RM programs \citep[e.g.,][]{bentz2006, derosa2018} are located around minima, while our RM campaign is situated on the rise and near the second outburst. \citet{kaspi1996} made an RM campaign in 1993, when NGC~4151 is in the rising period of the first outburst. Section~\ref{sec:comparison} will compare these RM results in details.

\subsection{Changing kinematics of BLR} 
NGC~4151 has been monitored by several RM campaigns in the past, some of which provided velocity-resolved results that allowed us to study its BLR properties. The velocity-delay map for \civ $\lambda$1549 was first measured in 1991, where the BLR structure of \civ\ is asymmetric. Also, the time lags are 2 days in the strong redshifted line wing, and less than 10 days in the weak blueshifted line wing, indicating that gas motion is infalling \citep{Ulrich1996}. Additionally, \citet{bentz2022} obtained the kinematics of BLR by modeling the RM data in 2005. And the motions of BLR are dominated by eccentric bound orbits, but 10\% of orbits are near circular motions. In 2012, the BLR structure of \hb\ was investigated, and the BLR was found to be in a virialized motion \citep{derosa2018}. However, when checking their result, we find that the redshifted velocity-bin lag is larger than the blueshifted velocity-bin lag at the same velocity, which reveals a signature of outflow.  In this campaign, we measure multiple emission-line velocity-resolved delays, but virial and infall signatures coexist. In general, the kinematics of BLR have changed for unknown reasons, possibly suggesting an evolution of the BLR. 

\subsection{Influence of Time Lags between UV and B Bands} \label{sec:uv/optical}
In 2016, a 69-day multi-band, covering X-ray, UV, and optical bands, monitoring campaign of NGC~4151 was carried out at {\it{Swift}} \citep{edelson2017}. We note that the time delay between $uvw2$ (with a central wavelength of 1928 \AA) and $b$ (with a central wavelength of 4395 \AA) bands is $0.83_{-0.34}^{+0.32}$ days, denoted by $\tau_{\rm b-uv}$. This value can be regarded as the size of radiation region of the $B$ band, which should not be ignored in the analysis of lags between the BLR and ionization source. To date, there is no clear evidence that the size of the $B$/optical band can change significantly. Therefore, we sum it with our measured $\tau_{\rm Mean}$ and write it as $\tau_{\rm Tot}$ after being corrected to the rest frame. Meanwhile, we employ $\tau_{\rm Tot}$, FWHM, and $f=1.3$ to compute $M_{\bullet}$, denoted as $M_{\bullet}^{'}$ in Table~\ref{table:width}. 

The rations of $\tau_{\rm Tot}$ to $\tau_{\rm Mean}$ increased by 0.108, 0.133,  0.146, 0.521, 1.815 in \ha, \hb, \hg, \hei, and \heii, respectively. Correspondingly, the black hole mass also has the same change. The UV/optical lag has a significant impact on the results of \hei\ and \heii, but a less impact on those of the Balmer lines. Moreover, considering the UV/optical lag would reduce the discrepancy in measuring the mass of the black hole using different emission lines. 

\subsection{Virial Test} \label{sec:vr}
\begin{figure*}[!ht]
\centering 
\includegraphics[width=0.49\textwidth]{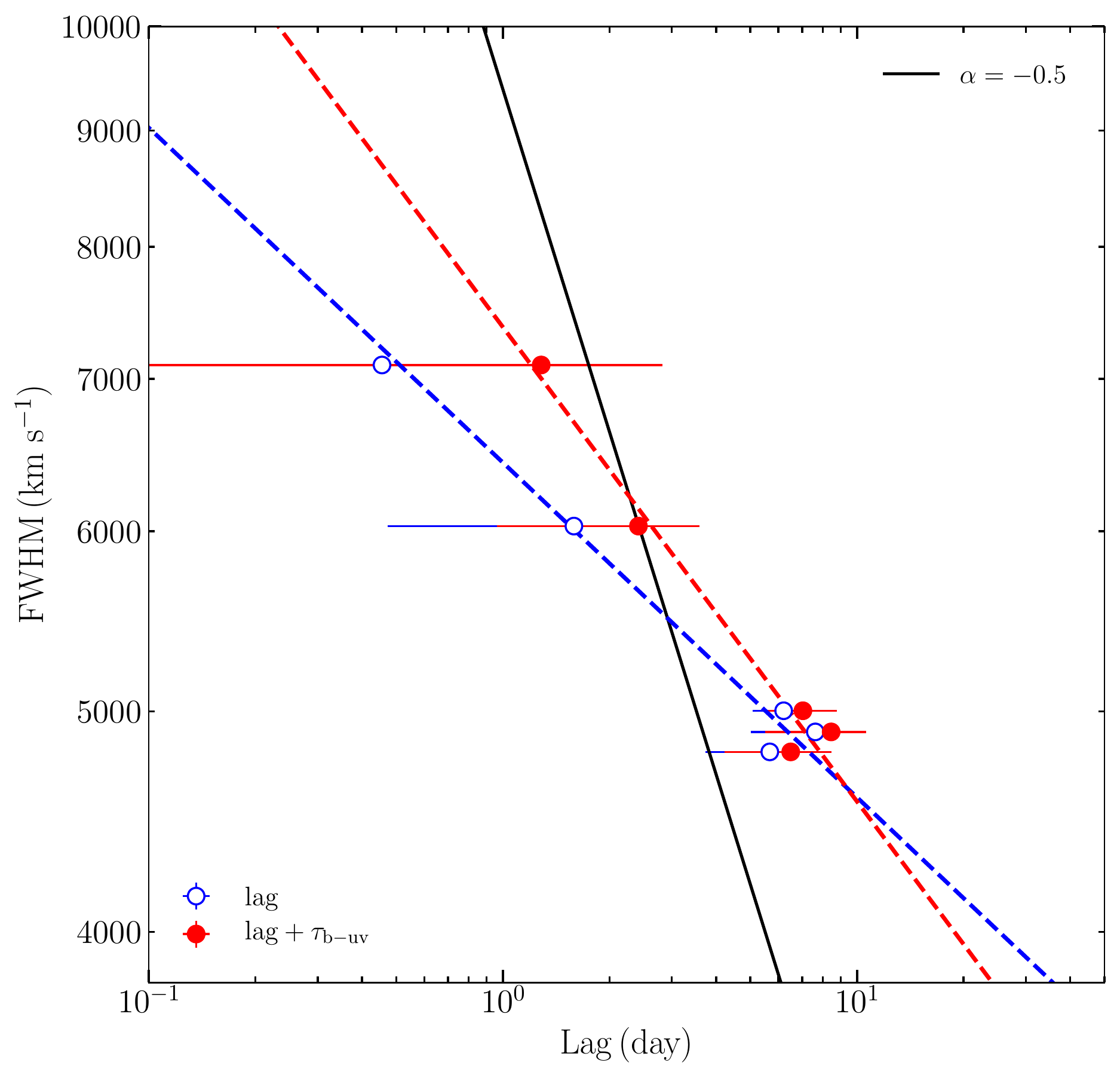}
\includegraphics[width=0.49\textwidth]{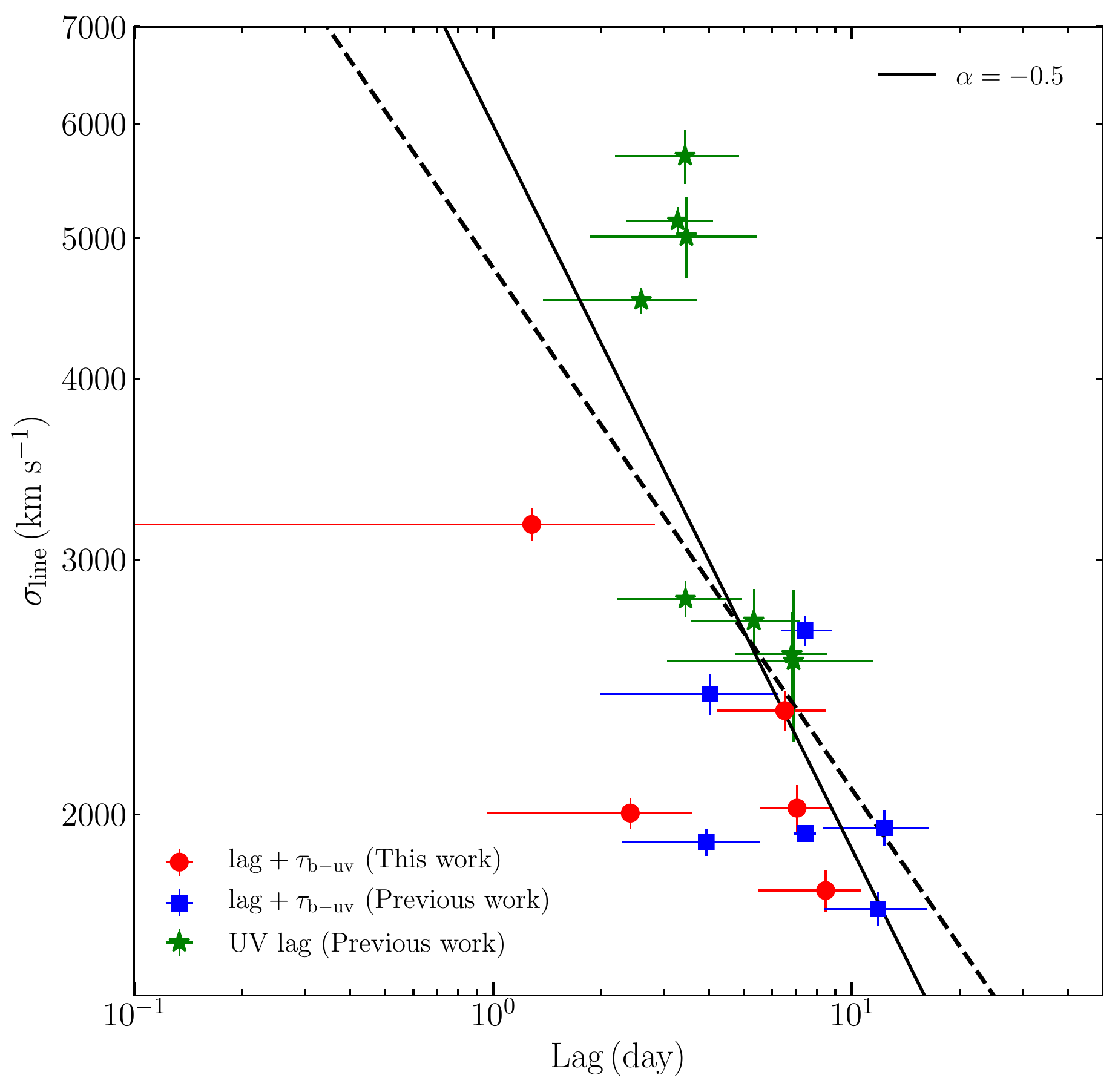}
\caption{Line width vs. time lag. The left panel: our observational results of FWHM vs. time lag. The blue data is $\tau_{\rm Mean}$, and the red data is $\tau_{\rm Mean}$ plus $\tau_{\rm b-uv}$. The blue and red dashed lines represent the best fit lines with slopes of $-0.15 \pm 0.04$ and $-0.21 \pm 0.06$, respectively. The right panel: rms $\sigma_{\rm line}$ vs. time lag for all the data. The black dashed line represents the best fit line with a slope of $-0.36\pm 0.10$. The black solid line in each panel is the best fit line using the virial relation with a slope of -0.5.}   
\label{fig:vr}
\end{figure*}

In order to investigate the motion of BLR in NGC~4151, we should check whether it is governed by the gravity of the black hole and obeys the virial relation ($\Delta V \propto \tau^{-0.5}$). This can be achieved by the following test. First, we investigate this relation through our estimates of time lags and velocities. Regression analysis of $\Delta V \propto \tau^{\alpha}$ is run for $\tau_{\rm Mean}$ and $\tau_{\rm Tot}$ with FWHMs measured from the mean spectrum. The best fittings give slopes of $\alpha=-0.15 \pm 0.04$ and $-0.21 \pm 0.06$ for the $\tau_{\rm Mean}$ and $\tau_{\rm Tot}$ data, respectively (see Figure~\ref{fig:vr}). The two best-fit results deviate from that of the virial relation with $\alpha=-0.5$. It is evident that the slope of applying $\tau_{\rm Tot}$ is closer to the virial relation but still deviates from it. This might be caused by too few data points. Therefore, we collected time lags and line widths from the UV emission-line measurements and other optical RM results. Our data and those collected data are listed in Table~\ref{table:comparison}. 

Considering that the lags of UV emission lines are not affected by accretion disk size because they are measured using UV continuum, we only add $\tau_{\rm b-uv}$ to the lags of optical lines for analysis. However, most previous results only provide emission-line widths of the rms spectrum. In comparison with them, we also present $\sigma_{\rm line}$ of the rms spectrum. Similar to the way for the mean spectrum described in Section~\ref{sec:bhmass}, the line widths and corresponding uncertainties are assessed using 1000 times bootstrapping rms spectrum. For an individual rms spectrum, the $\sigma_{\rm line}$ of each line is directly measured after subtracting the underlying continuum via a straight line. The straight line is fitted using two continuum windows on both sides of the emission line. The median and standard deviation of the $\sigma_{\rm line}$ distribution are taken as the final value of $\sigma_{\rm line}$ and associated error, respectively, while subtracting the instrumental broadening (see Table~\ref{table:comparison}). However, it should be stressed that the narrow-line residuals also appear in the rms spectrum, which will cause uncertainty in the estimate of $\sigma_{\rm line}$. We fit all the data of $\sigma_{\rm line}$ and time lag in Table~\ref{table:comparison}, and obtain a slope of $-0.36 \pm 0.10$ (see Figure~\ref{fig:vr}), which is marginally consistent with the expected value of $\alpha=-0.5$ within the error bar. However, it should be noted that there is also sizable scatter. Additionally, the data points are basically distributed around the virial relation of $\Delta V \propto \tau^{-0.5}$, implying that the BLR of NGC~4151 is basically in a virialized motion. 

\begin{deluxetable}{lccccccccc}[!ht]
\tablewidth{\textwidth}
\tabletypesize{\scriptsize}
\tablecaption{Rest-frame Lags, rms $\sigma_{\rm line}$ and Black Hole Masses \label{table:comparison}}
\tablewidth{\textwidth}
\tablehead{
	\multirow{2}{*}{Method} &
	\multicolumn{1}{c}{Lag}&
	\multicolumn{1}{c}{rms $\sigma_{\rm line}$}&
	\multicolumn{1}{c}{$M_{\bullet}~^d$}&
	\multirow{2}{*}{Reference}\\
	\colhead{}&
	\multicolumn{1}{c}{(day)}&
		\multicolumn{1}{c}{($\kms$)}&
	\multicolumn{1}{c}{($\times 10^7 M_{\odot}$)}
		}
\startdata   
\ha\ RM & $7.63_{-2.62}^{+1.85}$ & $1772\pm 59~^b$ & $2.94_{-1.03}^{+0.74}$ & This work \\ 
\hb\ RM & $6.21_{-1.13}^{+1.41}$ & $2020\pm 76~^b$ & $3.11_{-0.61}^{+0.75}$ & This work \\ 
\hg\ RM & $5.67_{-1.94}^{+1.64}$ & $2359\pm 73~^b$ & $3.88_{-1.35}^{+1.15}$ & This work \\ 
\hei\ RM & $1.59_{-1.11}^{+0.86}$ & $2004\pm 48~^b$ & $0.78_{-0.55}^{+0.43}$ & This work \\ 
\heii\ RM & $0.46_{-1.06}^{+1.22}$ & $3172\pm 82~^b$ & $0.56_{-1.31}^{+1.51}$ & This work \\
\ha\ RM & $11.00_{-3.10}^{+4.10}$ & $1721\pm 47~^b$ & $4.01_{-1.15}^{+1.51}$ & 1, 3\\ 
\hb\ RM & $11.50_{-3.70}^{+3.70}$ & $1958\pm 56~^b$ & $5.42_{-1.77}^{+1.77}$ & 1, 3\\ 
\ha\ RM & $3.20_{-1.70}^{+1.90}$ & $2422\pm 79$ & $2.31_{-1.24}^{+1.38}$ & 2, 3\\ 
\hb\ RM & $3.10_{-1.30}^{+1.30}$ & $1914\pm 42$ & $1.40_{-0.59}^{+0.59}$ & 2, 3\\ 
\hb\ RM & $6.57_{-0.72}^{+1.12}$ & $2680\pm 64$ & $5.80_{-0.69}^{+1.03}$ & 4\\ 
\hb\ RM & $6.59_{-0.21}^{+0.19}$ & $1940\pm 22$ & $3.05_{-0.12}^{+0.11}$ & 5\\ 
\civ\ RM & $3.43_{-1.24}^{+1.42}$ & $5698\pm 245$ & $13.69_{-5.09}^{+5.79}$ & 6, 8 \\ 
\heii\ RM $^a$ & $3.46_{-1.60}^{+1.96}$ & $5013\pm 323$ & $10.69_{-5.13}^{+6.21}$ & 6, 8 \\ 
\ciii\ RM & $6.88_{-3.82}^{+4.56}$ & $2553\pm 307$ & $5.51_{-3.34}^{+3.89}$ & 6, 8 \\ 
\mgii\ RM & $6.81_{-2.09}^{+1.73}$ & $2581\pm 179$ & $5.58_{-1.88}^{+1.61}$ & 6, 8 \\ 
\civ\ RM & $3.27_{-0.91}^{+0.83}$ & $5140\pm 113~^c$ & $10.62_{-2.99}^{+2.74}$ & 7, 8 \\ 
\heii\ RM $^a$ & $2.59_{-1.21}^{+1.10}$ & $4530\pm 92$ & $6.54_{-3.06}^{+2.79}$ & 7, 8 \\ 
\ciii\ RM & $3.44_{-1.22}^{+1.51}$ & $2817\pm 81$ & $3.36_{-1.21}^{+1.49}$ & 7, 8 \\ 
\mgii\ RM & $5.33_{-1.76}^{+1.86}$ & $2721\pm 141$ & $4.85_{-1.68}^{+1.77}$ & 7, 8 \\ 
Gas dynamics &  ---   &   ---  &$3.6_{-2.6}^{+0.9}$ & 9, 11 \\ 
Stellar dynamics &   ---   &   --- & $4.27_{-1.31}^{+1.31}$ & 10, 11 \\ 
Stellar dynamics &  ---   &   ---  & $0.25-3.0$ & 11 \\ 
RM modeling &  ---   &   ---  & $1.66_{-0.34}^{+0.48}$ & 12 \\ 
\enddata
\tablecomments{References: (1) \citet{maoz1991}, (2) \citet{kaspi1996}, (3) \citet{peterson2004}, (4) \citet{bentz2006}, (5) \citet{derosa2018}, (6) \citet{Clavel1990}, (7) \citet{Ulrich1996}, (8) \citet{Metzroth2006}, (9) \citet{hicks2008}, (10) \citet{onken2014}, (11) \citet{roberts2021}, (12) \citet{bentz2022}. \\
$^a$ Here \heii\ refers to the \heii\,$\lambda$1640 line.\\
$^b$ The value is uncertain, since the rms spectrum incorporate the narrow line residuals, and bring unconvinced line width measurements.\\
$^c$ The value is also unsure, because that \civ\ line is strongly self-absorbed.\\ 
$^d$ The values of $M_{\bullet}$ measured by RM are corrected by $f=6.3$, which is much larger than $f=1.3$ because $f=6.3$ corresponds to $\sigma_{\rm line}$ rather than FWHM.}
\end{deluxetable}

\subsection{Comparison with Previous Results} \label{sec:comparison}
We mainly compare our measurements of broad-line lags and black hole mass of NGC~4151 with other published results. Historically, multiple RM campaigns for NGC~4151 had been carried out, and had measured the relevant time lags of optical and/or UV emission lines. Here, we separately introduce their results. In the RM monitoring of 1988, the time lag was $9\pm 2$ days in both \ha\ and \hb\ \citep{maoz1991}. \citet{kaspi1996} measured time lags of 0--3 days for \ha\ and \hb\ with respect to continuum variations in 1993. Later, the above two data sets are reanalyzed by \citet{peterson2004}. Then, \citet{bentz2006} and \citet{derosa2018} performed well-sampled RM campaigns in 2005 and 2012, respectively, and obtained $6.57_{-0.76}^{+1.12}$ and $6.59_{-0.21}^{+0.19}$ days of \hb\ lags at the rest frame. In our work, the time lags of \ha\ and \hb\ are basically consistent with the above results of others. Additionally, \citet{Clavel1990} and \citet{Ulrich1996} performed UV RM programs. \citet{Metzroth2006} reanalyzed their data and obtained the corresponding lags and line widths.

On the other hand, the black hole mass of NGC~4151 has been measured by different methods, for instance, the gas dynamical modeling \citep{hicks2008}, the stellar dynamical modeling \citep{onken2014, roberts2021}, the RM method \citep{peterson2004, bentz2006, Metzroth2006, derosa2018}, and the RM data modeling \citep{bentz2022}. Thus, we can compare these measurements to check the reliability of different methods. The black hole masses measured by these methods are listed in Table~\ref{table:comparison}. Here, the dynamical modeling results were unified by \citet{roberts2021} according to the Cepheid distance of $15.8 \pm 0.4$~Mpc \citep{yuan2020}. Because the virial coefficients used in different RM projects are different, we adopt $f =6.3$ in \citet{ho2014} to adjust the RM results. Additionally, for consistency, we also give the values of $M_{\bullet}$ estimated with rms $\sigma_{\rm line}$ measured in Section~\ref{sec:vr} (see Table~\ref{table:comparison}). It's worth noting that the black hole masses determined by different methods are consistent within errors. However, the results for the UV lines are overall larger than others, even an order of magnitude.

\section{Conclusion} \label{sec:conclusion}
Here, we report a new RM campaign on NGC~4151 and the detailed results are summarized below. 
\begin{enumerate}
\item We measure time delays between multiple broad emission lines and continuum simultaneously, to be $7.63_{-2.62}^{+1.85}$, $6.21_{-1.13}^{+1.41}$, $5.67_{-1.94}^{+1.65}$, $1.59_{-1.11}^{+0.86}$, and $0.46_{-1.06}^{+1.22}$ days for the broad \ha, \hb, \hg, \hei, and \heii\ lines, respectively. These lags satisfy $\tau_{\rm H\alpha} > \tau_{\rm H\beta} > \tau_{\rm H\gamma} > \tau_{\rm He\ I} >\tau_{\rm He\ II}$, which are radial stratification and may be affected by optical depths and ionization energy. Besides that, the time lag ratios of \ha, \hb, \hg, \hei, and \heii\ relative to \hb\ are $1.23 : 1.00 : 0.91 : 0.26 : 0.07$. If considering the time lag of the optical band relative to the UV band, it would bring more variation in the time lags of \hei\ and \heii. 

\item It's the first time for NGC~4151 to explore velocity-resolved time lags of multiple optical emission lines. The velocity-resolved structures of \ha, \hb, and \hei\ are consistent with each other, the lags of the blue wing are overall larger than those of the red wing, and the lags of the line core are larger than those of both wings, indicating that virial and infalling motions coexist. \hg\ may be affected by \oiii\,$\lambda$4363, and \heii\ is weak and likely contaminated by \feii, causing less reliable results for \hg\ and \heii. In comparison with previous velocity-resolved structures, we note that the kinematics of the BLR changes from inflow to virialization with outflow, and then to virial and inflow coexistence signature, indicating the evolutionary kinematics of the BLR.

\item Combining our measurements and the collected RM results including time delays and velocities, we verify that all the observational data of time lag and velocity for NGC~4151 basically follows a virial relation of $\Delta V \propto \tau^{-0.5}$.

\item We measure the black hole mass of NGC~4151 based on the FWHM of the mean spectrum, the time lag and $f =1.3$, and the value is $ 3.94_{-0.72}^{+0.90} \times 10^7 \sunm$ for \hb. Furthermore, comparing our measurements with previous results, we find that they are consistent within errors. Using the black hole mass measured by \hb, we calculate the dimensionless accretion rate to be $0.02_{-0.01}^{+0.01}$, indicating NGC~4151 is in a sub-Eddington accretion state.
\end{enumerate}

NGC~4151 is one of the best-studied AGNs with multiple RM observations. Our result of the time lag of \hb\ is consistent with \citet{derosa2018}, but the velocity-resolved results are quite different, which hints the evolution of kinematics of the BLR. This campaign of NGC~4151 is located at the rising phase of the second outburst, and it is worth continuously monitoring NGC~4151 to explore the origin of variable kinematics.

\vspace{5mm}
We thank the referee for useful comments that improved the manuscript. We are grateful to Zi-Xu Yang, Jun-Rong Liu, and Zhu-Heng Yao for their time and effort in our observations. This work is supported by the National Key R\&D Program of China with No. 2021YFA1600404, the National Natural Science Foundation of China (NSFC; grants No. 11991051, 12103041, 12073068), the joint fund of Astronomy of the NSFC and the CAS (grant No. U1931131), and the science research grants from the China Manned Space Project with NO. CMS-CSST-2021-A06.

We acknowledge the support of the staff of the Lijiang 2.4m telescope. Funding for the telescope has been provided by Chinese Academy of Sciences and the People's Government of Yunnan Province.
The Zwicky Transient Facility Collaboration is supported by U.S. National Science Foundation through the Mid-Scale Innovations Program (MSIP). This research has made use of the NASA/IPAC Extragalactic Database (NED) which is operated by the Jet Propulsion Laboratory, California Institute of Technology, under contract with the National Aeronautics and Space.

\vspace{5mm}

\facilities{Lijiang: 2.4 m}

\software{IRAF \citep{tody1986}, DASpec (\url{https://github.com/PuDu-Astro/DASpec}), PyCALI \citep{li2014L}, MICA \citep{li2016}, JAVELIN\citep{zu2011}
}

\appendix
\section{Velocity-resolved Light Curves} \label{sec:appendix}
\begin{figure*}[!ht]
\centering 
\includegraphics[width=0.48\textwidth]{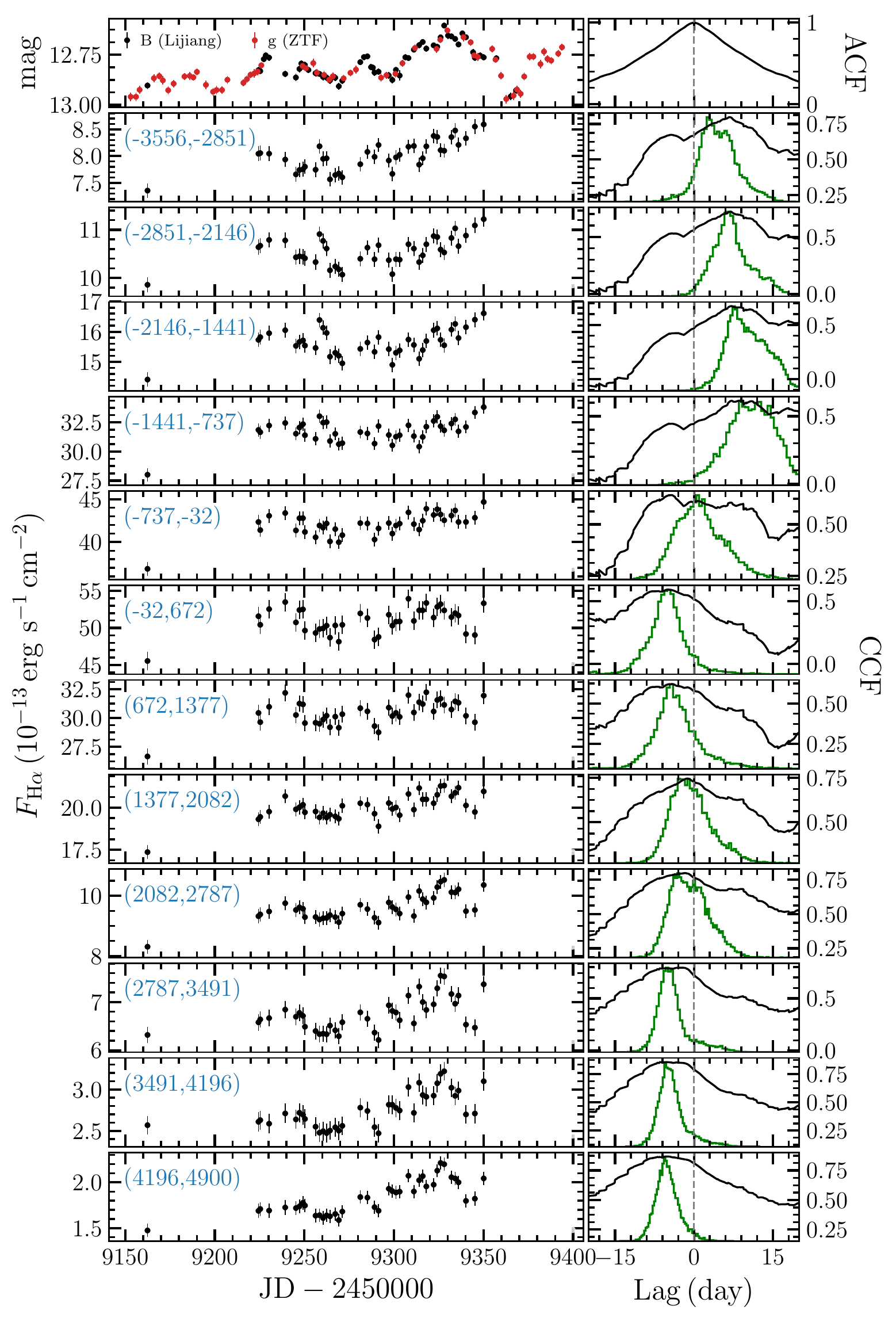}
\includegraphics[width=0.48\textwidth]{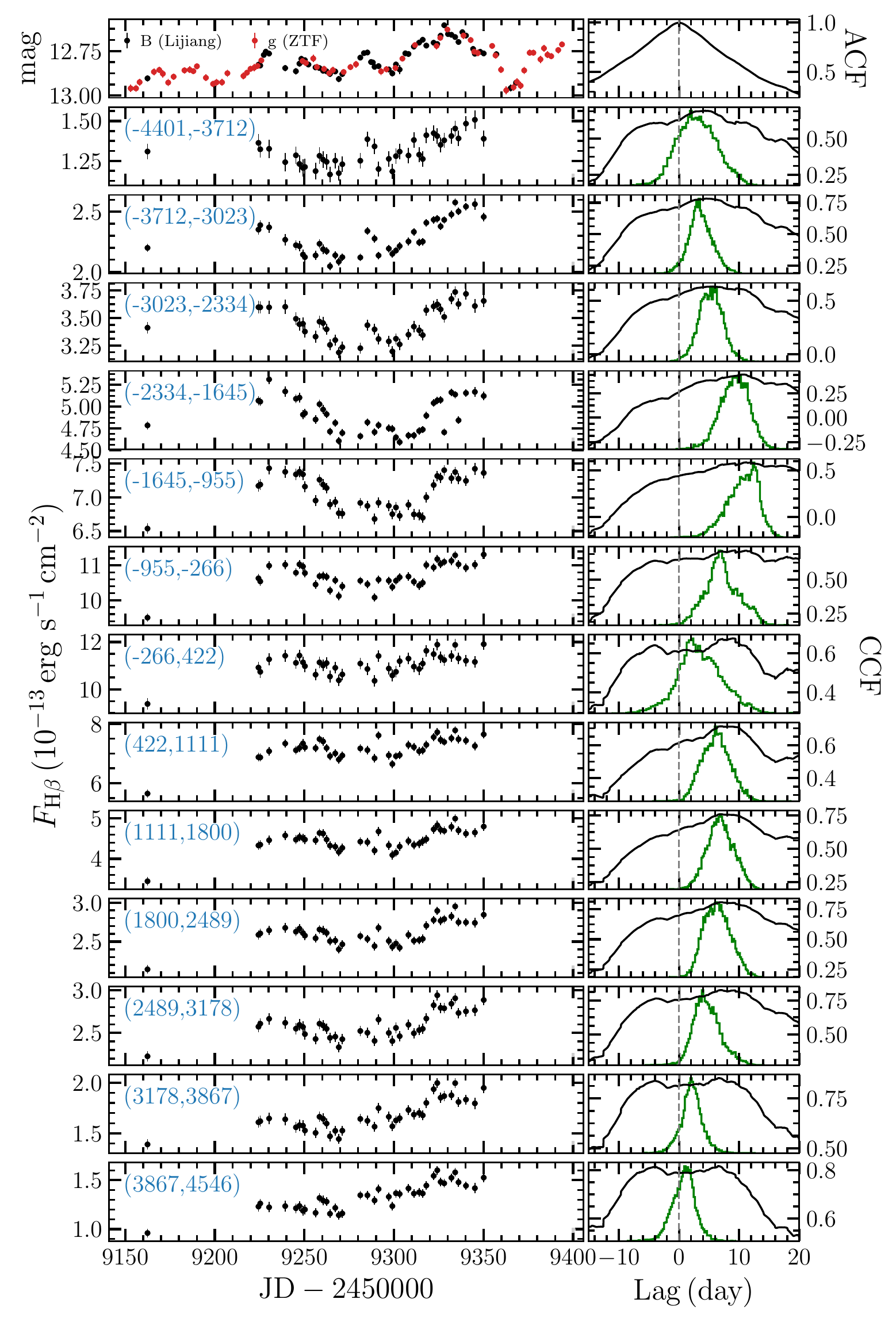}
\caption{The \ha\ (left) and \hb\ (right) results. The left panels are photometric and each bin light curves, where the photometry is $g$ (ZTF) merged with $B$ (Lijiang) results. The right panels display ACF of photometry, the cross-correlation results between each bin and photometry, including the ICCFs (black) and CCCDs (green).} 
\label{fig:velocitylc}
\end{figure*}

\begin{figure*}[!ht]
\centering 
\includegraphics[width=0.48\textwidth]{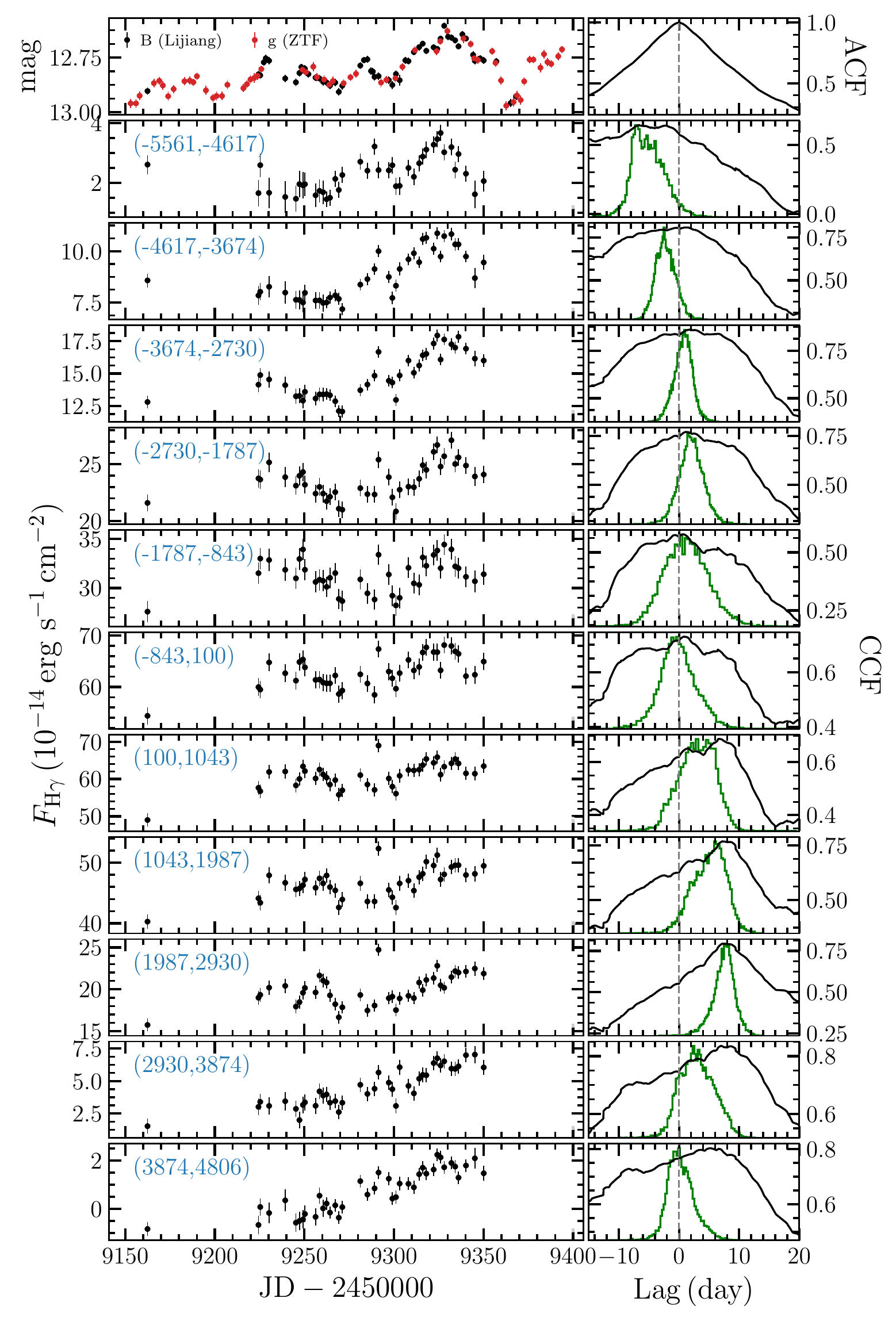}
\includegraphics[width=0.48\textwidth]{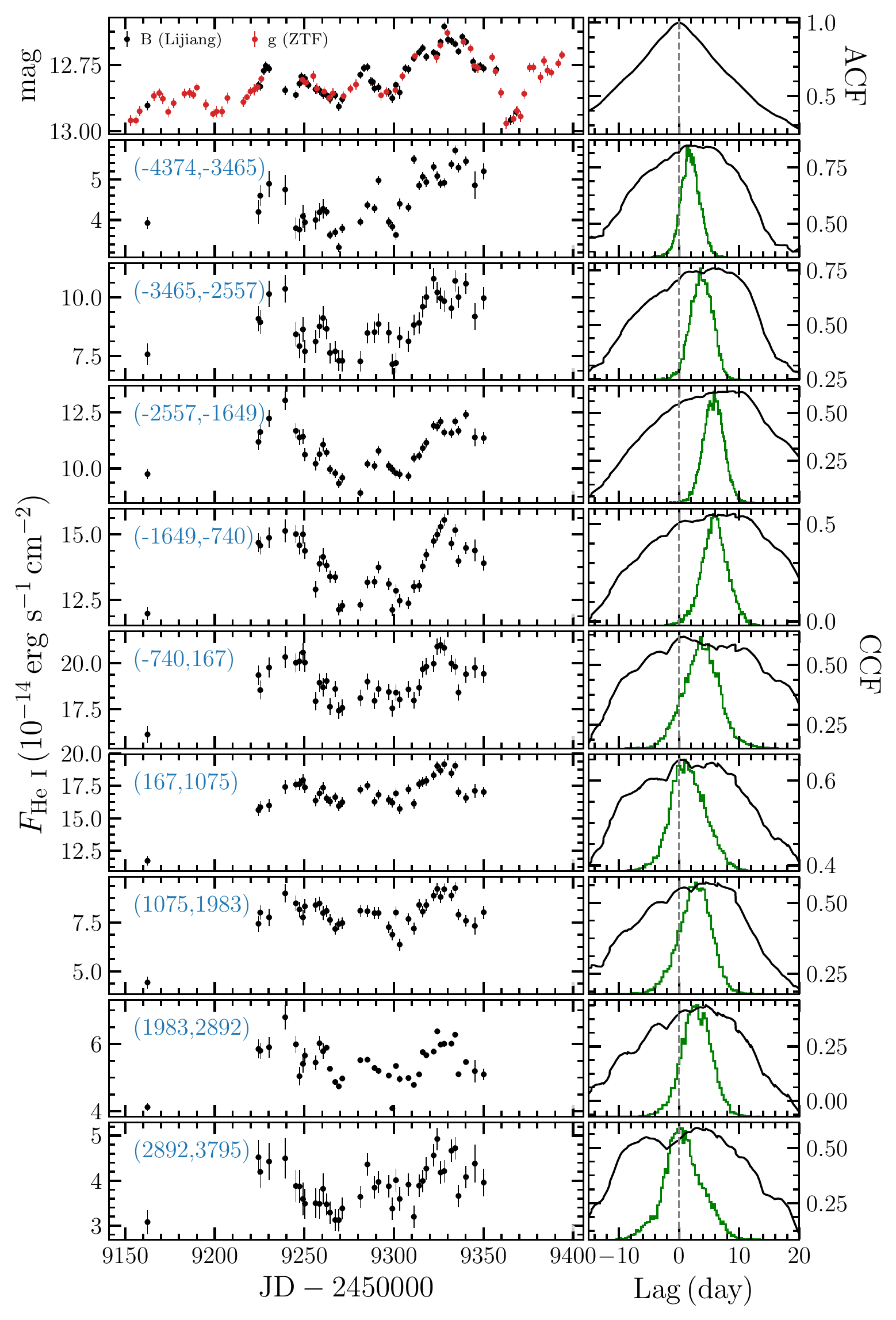}
\caption{Same as Figure~\ref{fig:velocitylc} but for the \hg\ (left) and \hei\ (right) results.}
\label{fig:velocitylc2}
\end{figure*}

\begin{figure*}[!ht]
\centering 
\includegraphics[width=0.48\textwidth]{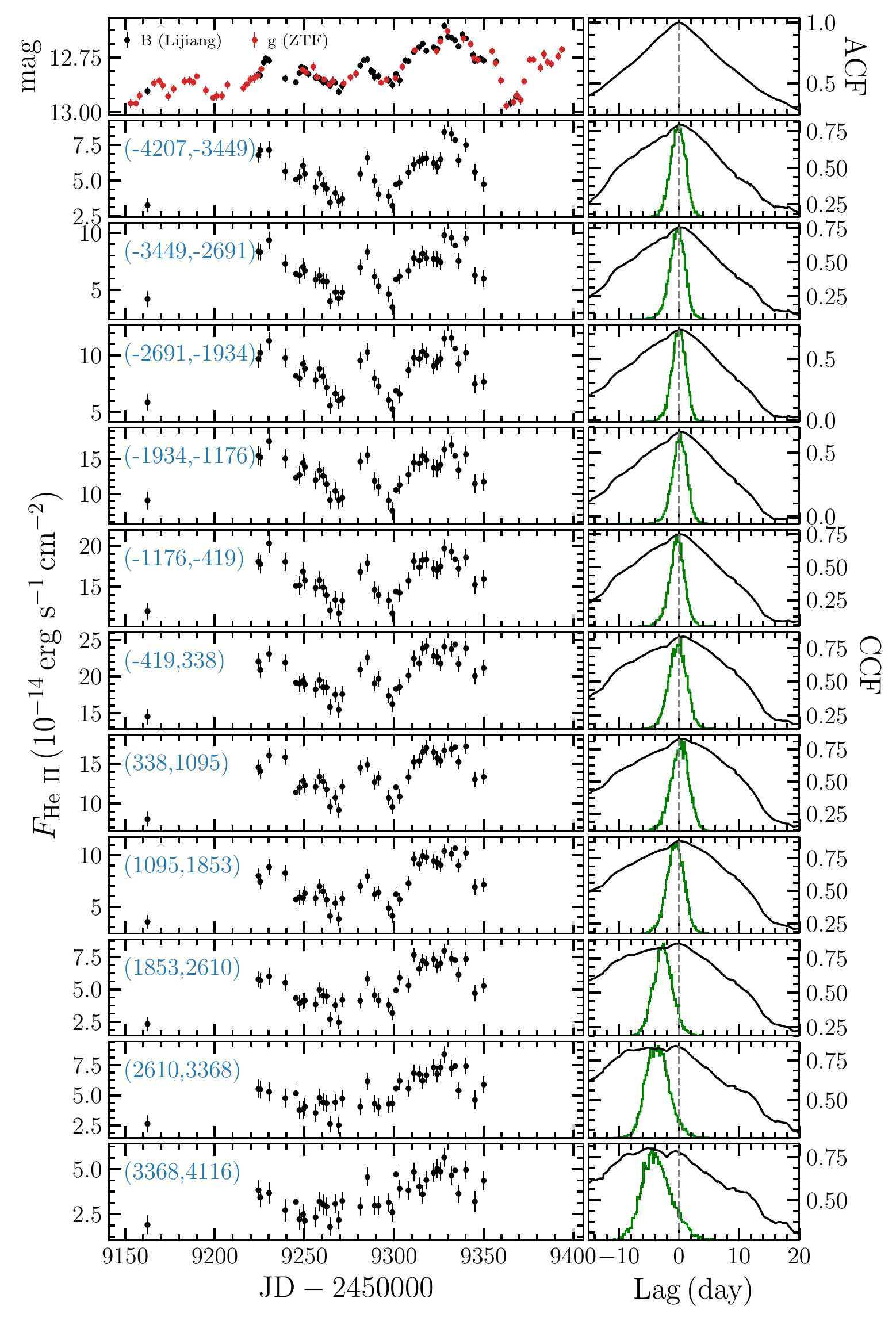}
\caption{Same as Figure~\ref{fig:velocitylc} but for the \heii\ results.}
\label{fig:velocitylc3}
\end{figure*}

Here are velocity-space bin light curves and cross-correlation results for \ha, \hb, \hg, \hei, and \heii, illustrated in Figure~\ref{fig:velocitylc}, \ref{fig:velocitylc2} and \ref{fig:velocitylc3}.

\end{document}